\documentclass[journal=jctcce,manuscript=article,layout=twocolumn]{achemso}

\usepackage[version=3]{mhchem} % Formula subscripts using \ce{}
\usepackage{amsmath}
\usepackage{mathtools}
\usepackage{graphicx} 
\usepackage{caption}
\usepackage{float}
\usepackage{subcaption}
\usepackage{physics}
\usepackage{fnpos}
\usepackage{amssymb}
\usepackage{charter}
\usepackage{xr}
\makeatletter
\newcommand*{\addFileDependency}[1]{% argument=file name and extension
  \typeout{(#1)}% latexmk will find this if $recorder=0 (however, in that case, it will ignore #1 if it is a .aux or .pdf file etc and it exists! if it doesn't exist, it will appear in the list of dependents regardless)
  \@addtofilelist{#1}% if you want it to appear in \listfiles, not really necessary and latexmk doesn't use this
  \IfFileExists{#1}{}{\typeout{No file #1.}}% latexmk will find this message if #1 doesn't exist (yet)
}
\makeatother

\newcommand*{\myexternaldocument}[1]{%
    \externaldocument{#1}%
    \addFileDependency{#1.tex}%
    \addFileDependency{#1.aux}%
}

\myexternaldocument{./SIjctc}
\usepackage{hyperref}
\DeclareUnicodeCharacter{03A3}{$\Sigma$}

\usepackage{xcolor}
\definecolor{delftblue}{RGB}{31, 48, 94}
\definecolor{cafenoir}{RGB}{75, 54, 33}
\definecolor{jet}{RGB}{52, 52, 52}
\definecolor{col1}{rgb}{0,0.466667,0.733333}
\definecolor{col2}{rgb}{0.2,0.733333,0.933333}
\definecolor{col3}{rgb}{0,0.6,0.533333}
\definecolor{col4}{rgb}{0.933333,0.466667,0.2}
\definecolor{col5}{rgb}{0.8,0.2,0.0666667}
\definecolor{col6}{rgb}{0.933333,0.2,0.466667}
\definecolor{col7}{rgb}{0.733333,0.733333,0.733333}
\definecolor{col8}{rgb}{0.392157,0.392157,0.392157}

\usepackage{siunitx}
\usepackage{commath}

\author{Kyle Acheson}
\affiliation[University of Edinburgh]
{EaStCHEM, School of Chemistry and Centre for Science at Extreme Conditions, University of Edinburgh, David Brewster Road, Edinburgh EH9 3FJ, United Kingdom}
\author{Adam Kirrander}
\email{Adam.Kirrander@chem.ox.ac.uk}
\phone{+(0)1865 275422}
\fax{+(0)1865 275400}
\affiliation[University of Oxford]
{Physical and Theoretical Chemistry Laboratory, Department of Chemistry, University of Oxford, South Parks Road, Oxford OX1 3QZ, United Kingdom}

\title{Robust inversion of time-resolved data via forward-optimisation in a trajectory basis}
\abbreviations{NMR, TRPES, UED, UXS, CEI, MCTDH, dd, SH, AIMS, AIMCE, dd-vMCG, TBFs, BAT, CHD, HT, IAM, SI, RAE, RMSE, PDF}

\begin{document}

\begin{tocentry}
\includegraphics{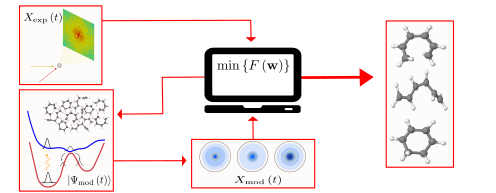}
A new inversion method for data from ultrafast experiments, mapping photochemical reactions.
%Schematic of the inversion of data from an experiment studying ultrafast photochemical dynamics. A model wavefunction $\ket{\Psi_{\mathrm{mod}}\left(t\right)}$ is adjusted until the predicted observables $X_{\mathrm{mod}}\left(t\right)$ are in optimal agreement with the experimental observables $X_{\mathrm{exp}}\left(t\right)$ via least squares minimisation of the target function $F\left(\mathbf{w}\right)$. The outcome is a detailed molecular model that agrees with experimental data.
\end{tocentry}


% ABSTRACT
\begin{abstract}
An inversion method for time-resolved data from ultrafast experiments is introduced, based on forward-optimisation in a trajectory basis. The method is applied to experimental data from x-ray scattering of the photochemical ring-opening reaction of 1,3-cyclohexadiene and electron diffraction of the photodissociation of \ce{CS2}. In each case, inversion yields a model that reproduces the experimental data, identifies the main dynamic motifs, and agrees with independent experimental observations. Notably, the method explicitly accounts for continuity constraints and is robust even for noisy data. 
\end{abstract}


\section{Introduction}

Inverse problems are important across science and engineering,\cite{Tanaka_inverse_1998,Razavy_inverse_2020,Tanaka_deep_2021} and have a long history at the borderline between physics and mathematics.\cite{Giraud2010} They concern the determination of key target characteristics from an observed set of outputs, for instance determining the shape of a vibrating membrane from the spectrum or reconstructing an object from a photograph. Important applications appear in medical imaging and material characterisation.\cite{Tanaka_inverse_1998,Arridge_optical_1999,Bertero_inverse_2006,Razavy_inverse_2020,Tanaka_deep_2021} In contrast to the forward, also known as the direct, problem, inverse problems are often mathematically ill-posed, meaning they may exhibit instability, be underdetermined, or lack unique solutions. These challenges must be overcome via new computational schemes, regularization techniques, objective functionals, and experimental procedures. 

For molecular structure determination the inverse problem is central. The structure of a molecule is generally underdetermined by the experimental data and the molecular model is optimised (bootstrapped/refined) subject to auxiliary constraints. Examples include the fitting of spectroscopic data to a Morse oscillator or Dunham coefficients in a diatomic,\cite{Herzberg1979} bond-distance, bond-angle, and torsional constraints during x-ray crystallography or NMR refinement,\cite{Als-Nielsen2} and the use of supplementary experimental or computational data in gas-phase electron diffraction structure determination.\cite{Hargittai} 

In recent years, ultrafast imaging of photoexcited molecules have developed rapidly, with experiments increasingly capable of tracking molecular dynamics on fundamental timescales, observing phenomena such as vibrations, bond breaking, or charge transfer.\cite{stankus_ultrafast_2019,gabalski_transient_2022,yong_determination_2021} The techniques include spectroscopy (e.g. TRPES),\cite{zewail_femtochemistry_2000,Stolow2004review,StolowFD2016,Penfold2015} ultrafast electron diffraction (UED),\cite{centurion_ultrafast_2016,ischenko_capturing_2017} ultrafast x-ray scattering (UXS),\cite{BudarzJPB2015,stankus_advances_2020} Coulomb explosion imaging (CEI),\cite{Rolles2023review} and others. Beyond fitting data to heuristic rate models, extracting a more detailed molecular model is often challenging. From the perspective of structure determination, for instance, a minimal structural model would consist of a sequence of structures fulfilling at least some degree of temporal continuity. A complete model would conceivably consist of the full temporal evolution of the molecular wavefunction with all associated time-dependent nuclear and electronic distributions and populations. 

Part of the issue is that most experiments only provide partial information of complex and multidimensional dynamics, which makes the complete problem severely underdetermined. One obvious strategy is to combine data from complementary experiments. For instance, spectra are sensitive to electronic populations while scattering experiments primarily, but not exclusively, detect the molecular geometry.\cite{Bressler2015,pemberton_1b_2015,Stankus2016,tudorovskaya_effects_2018,yong_determination_2021} Another important strategy is comparison to theoretical results. Least but not last, inversion algorithms that account for at least some of the physics stand to play an increasingly important role.

The challenges involved in inversion in this context are so significant that the default option is to rely on comparison between theory and experiment only. The more reliable variety of this approach makes the comparison in the observable space, with observables forward mapped (i.e.\ predicted) from simulations. However, such comparison only provides a qualitative understanding of the data and leaves little recourse if agreement between experiment and theory is not achieved. Inversion algorithms capable of providing a rapid, first-order, assessment of the observed dynamics would carry great value for experimental progress, even if by simply determining the range of molecular models that are, in principle, commensurate with the observations.

Therefore, efforts to tackle the inverse problem for ultrafast dynamics are intensifying. One recently developed approach used for the interpretation of ultrafast x-ray scattering data samples a large pool of randomly generated molecular structures to find the best fit,\cite{yong_scattering_2019, YongFD2021, yong_determination_2021} and genetic algorithms have been proposed to invert electron diffraction data by exploring molecular structure via consecutive {\it{in silico}} mutations.\cite{habershon_determining_2006,hensley_imaging_2012,yang_ultrafast_2013,yang_imaging_2014,yang_reconstruction_2014} These approaches include phase retrieval algorithms for the structural determination, but this requires that the molecules are at least partially aligned.\cite{saldin_new_2011, saldin_reconstruction_2010,saldin_reconstructing_2011,shneerson_crystallography_2008,fung_structure_2009,starodub_single-particle_2012} The preceding approaches are applicable to both static and dynamic data, but methods specifically aimed at dynamic data have also been attempted. This includes machine learning approaches that use a variational recurrent neural network trained on temporally correlated frames \cite{Asenov2020} and an approach that systematically perturbs the molecular structure at each step, starting with the well-known initial structure at time zero.\cite{ishikawa_direct_2015}

In this paper, we present a detailed discussion of an approach for modelling time-dependent data which explicitly includes the time-evolution, thereby ensuring that continuity, inclusion of known initial or final structures and any other known physics. It proceeds by optimising the weights of semiclassical trajectories from quantum molecular dynamics simulations against experimental data. In essence, this biases the simulations towards experimental observations. The approach is general, and provides a platform for merging data from several complementary experiments, with initial focus on the analysis of experimental data from ultrafast x-ray and electron scattering experiments.\cite{minitti_toward_2014, minitti_imaging_2015, razmus_multichannel_2022} The aim of the paper is to provide the first unified and general presentation of this methodology, and to critically evaluate its performance, establish best-practice, and explore avenues for improving the methodology further.

\section{\label{sec:Theory}Theory}

\subsection{Forward optimisation}
The time evolution of any observable $X(t)$ can be calculated from the molecular wavefunction $\ket{\Psi(t)}$ via a forward mapping $M$,
\begin{equation} \label{eq:forward}
    \ket{\Psi\left(t\right)} \; \xrightarrow{M_j} \; X^j(t),
\end{equation}
where the index $j$ identifies the type of measurement, which could be anything from photoelectron spectra to electron diffraction or Coulomb explosion imaging. The observable $X^j(t)$ may be resolved with respect to several implicit variables, for instance photoelectron kinetic energy and angular distribution. In contrast to the inverse problem, the forward mapping is mathematically well-conditioned and does not suffer stability or underdetermination issues. We therefore proceed to tackle the inverse as a forward optimisation problem,\cite{minitti_toward_2014, minitti_imaging_2015, razmus_multichannel_2022} where our goal is to find a model molecular wavefunction $\ket{\Psi_{\mathrm{mod}}\left(t\right)}$ that yields predicted observables $\tilde{X}_\mathrm{mod}^{j}(t)$ that reproduce the experimental observables $X_\mathrm{exp}^j(t)$. The calculation of the observable consists of two steps, 
\begin{equation} \label{eq:predicted-data}
    \ket{\Psi_{\mathrm{mod}}\left(t\right)} \; \xrightarrow{M_j} \; X_\mathrm{mod}^{j}(t) \; \xrightarrow{S_j} \; \tilde{X}_\mathrm{mod}^{j}(t),
\end{equation}
where the first step is the forward mapping $M_j$ from eq~\ref{eq:forward}, which produces the theoretically predicted signal, and a second {\it{apparatus mapping}} step $S_j$ which replicates the effect of the measurement apparatus on the data, for instance due to limited time resolution. Such distortions are unavoidable despite that the experimental data $X_\mathrm{exp}^j(t)$ will have been through extensive preprocessing to remove known artefacts and distortions. 

The optimisation proceeds by modifying the model wavefunction until close agreement between the experimental and predicted data is achieved, as measured by the target function $F$,
\begin{equation}\label{eq:target_function}
    F = \sum_{j} \alpha_j \int \left|X_{\mathrm{exp}}^j(t)- \tilde{X}_{\mathrm{mod}}^j(t)\right|^2\;dt,
\end{equation}
where the index $j$ runs over the different types of experiments included. In practice it is common for the integration over time $t$ to be replaced by a summation over a temporal grid $\{t_i\}$. The $\alpha_j$ is a regularisation factor that must be included when data from several different types of experiments is considered. The factor is determined from the numerical profile of each of the different data sets, and scales data so that it can be combined in a balanced manner. 

The reminder of the paper is organised such that Section \ref{sec:reference-wf} presents the reference molecular wavefunction, Section \ref{sec:model-wf} presents the parameterised model wavefunction and the target function that result from the model function, and Section \ref{sec:scattering_theory} discusses the forward mapping with an emphasis on scattering experiments. In Section \ref{sec:comp}, the range of applicable numerical optimisation techniques are discussed, as well as the $S_j$ mapping to match the predicted observables to the experiment. The preprocessing of the experimental data is touched upon in Section~\ref{sec:methods} on data treatment and in the \textit{Supplementary Information} (SI). In the Results, Section \ref{sec:results}, two applications to recent ultrafast electron diffraction and ultrafast x-ray scattering data are examined in detail and the convergence and the resulting interpretation are assessed.

\subsection{Molecular wavefunction} \label{sec:reference-wf}
The time evolution of the molecular wavefunction $\ket{\Psi(t)}$ is governed by the time-dependent Schr\"odinger equation,
\begin{equation}\label{eq:tdse}
    i\hbar\frac{\partial}{\partial t}\ket{\Psi\left(t,\mathbf{r},\mathbf{R}\right)} = \hat{H}\ket{\Psi\left(t,\mathbf{r},\mathbf{R}\right)},
\end{equation}
where $\hat{H}$ is the molecular Hamiltonian and $\mathbf{r}$ and $\mathbf{R}$ the electronic and nuclear coordinates. The molecular wavefunction can be expanded in the Born-Huang form as,\cite{born_dynamical_1955}
\begin{equation}\label{eq:born_huang}
    \ket{\Psi\left(t\right)} = \sum_k^{\infty} \ket{\tilde{\chi}_k\left(t,\mathbf{R}\right)}\ket{\psi_k\left(\mathbf{r};\mathbf{R}\right)},
\end{equation}
where $\ket{\tilde{\chi}_k\left(t,\mathbf{R}\right)}$ are time-dependent nuclear wavepackets which propagate on electronic eigenstates $\ket{\psi_k\left(\mathbf{r};\mathbf{R}\right)}$ that depend parametrically on the nuclear coordinates $\mathbf{R}$.\footnote{In the following, $\mathbf{r}$ and $\mathbf{R}$ will be dropped from the equations.} In practice, the expansion in eq~\ref{eq:born_huang} is truncated to include only the $N_s$ electronic states visited during the dynamics of interest. 

A wide range of numerical techniques to solve eq~\ref{eq:tdse} exist. Accurate methods such as numerical grid propagators\cite{Feit1982,Nyman2013} and multiconfigurational time-dependent Hartree (MCTDH)\cite{LeticiaRoland_Ch12} require precalculated potential energy surfaces which are not feasible for most molecules of interest. An alternative is direct dynamics (dd) methods that expand the molecular wavefunction by classical or semiclassical trajectories. Examples include surface hopping (SH), {\it{ab initio}} multiple spawning (AIMS), {\it{ab initio}} multiconfigurational Ehrenfest (AIMCE), and direct dynamics variational multiconfigurational Gaussians (dd-vMCG).\cite{LeticiaRoland_Ch13,LeticiaRoland_Ch14,LeticiaRoland_Ch15,LeticiaRoland_Ch16} In a general form, the direct dynamics wavefunction $\ket{\Psi^\mathrm{dd}\left(t\right)}$ can be expressed as,
\begin{equation}\label{eq:tbf_approx}
    \ket{\Psi^\mathrm{dd}\left(t\right)} = \sum_{n=1}^{N_{\mathrm{TBF}}} c_{n}\left(t\right) \ket{\psi_{n} \left(t\right)},
\end{equation}
with $c_{n}\left(t\right)$ the expansion coefficient for each of the $N_{\mathrm{TBF}}$ trajectory basis functions (TBFs) given by,  
\begin{equation}\label{eq:tbf_product_ansatz}
    \ket{\psi_{n}\left(t\right)} = \left(\sum_{k=1}^{N_s} a^{n}_{k}\left(t\right) \ket{\psi_k}\right) \ket{\chi_n(t)},
\end{equation}
where the parenthesis contains the electronic states $\ket{\psi_k}$ and their populations $|a^{n}_{k}\left(t\right)|^2$. The nuclear basis functions $\ket{\chi_n(t)}$ follow phase-space trajectories $(\mathbf{R}_n(t),\mathbf{P}_n(t))$ where $\mathbf{R}_n(t)$ and $\mathbf{P}_n(t)$ are the nuclear positions and momenta, respectively. The equations of motion, which govern the trajectories, populations, and auxiliary coefficients such as for instance wavepacket width coefficients, are different for each method. Thus, the details of each specific dd wavefunction will vary. For instance, the nuclear wavepacket is a Gaussian\cite{Shalashilin2017} in all methods except SH where it is a $\delta$-function, for SH and AIMS only one electronic state is occupied by each TBF at any given time, while Ehrenfest methods such as AIMCE occupy several states simultaneously, in AIMS (AIMCE) spawning (cloning\cite{Makhov2014}) increases the number of TBFs as the simulation progresses, {\it{etc}}.

\subsection{Model wavefunction} \label{sec:model-wf}
We adapt the wavefunction in eq~\ref{eq:tbf_approx} as our parameterised model wavefunction by rescaling the expansion of the TBFs. In effect, the TBFs can be thought of physically reasonable constraints on a system far from equilibrium, which in addition automatically fulfill continuity requirements. The resulting model function is given by,
\begin{equation} \label{eq:model-wf}
    \ket{\Psi_{\mathrm{mod}}\left(t\right)} = \sum_n w_{n}\left(t\right) c_n \left(t\right) \ket{\psi_n \left(t\right)},
\end{equation}
where $w_n\left(t\right)$ are the weights for the TBFs which are adjusted to bias the theoretical model towards the experiment, subject to a normalisation condition $1=\braket{\Psi_{\mathrm{mod}}\left(t\right)}{\Psi_{\mathrm{mod}}\left(t\right)}$. 

If observables are taken to only depend on the nuclear coordinates and are calculated in the diagonal zeroth-order bracket-averaged Taylor expansion (BAT) approximation (see ref~\cite{kirrander_ultrafast_2016}), as we will do here, the coefficients $w_n(t)$ and $c_n(t)$ can be taken to be time-independent with the normalisation straightforwardly given by $\sum_n w_{n} = 1$ for $c_n\equiv1$. For SH wavefunctions, this is always the case. The target function in eq~\ref{eq:target_function} then becomes a function of the time-independent weights $\mathbf{w}=(w_1\ldots w_{N_\mathrm{TBF}})$.

%In the case of e.g.\ AIMCE, this gives a normalisation constant,
%\begin{equation}\label{eqn:wf_norm}
%    N\left(t\right) = \left({\sum_n  |w_{n}\left(t\right)|^2 \left(C_{n}^{*}\left(t\right) C_{n}\left(t\right)\right) \Delta_{n^{\prime}n}\left(t\right)}\right)^{-\frac{1}{2}} 
%\end{equation}
% \mbox{SGBF/ MGBF}\\ \left( \sum_n |w_{n}\left(t\right)|^{2} \right)^{-\frac{1}{2}} 
%with 
%\begin{equation}
%    \Delta_{n^{\prime}n}\left(t\right) = \left( \sum_{\beta=1}^{N_s} a_{\beta n'}^{*}\left(t\right) a_{\beta n}\left(t\right) \right) \bra{\chi_{n^{\prime}}\left(t\right)}\ket{\chi_{n}\left(t\right)},
%\end{equation}
%where $\bra{\chi_{n^{\prime}}\left(t\right)}\ket{\chi_{n}\left(t\right)}$ is the overlap of the two nuclear basis functions.\cite{Shalashilin2017} For SH, setting the coefficients $c_n(t)\equiv1$ gives the normalisation condition as $\sum_n w_{n}\left(t\right) = 1$. 

As a final aside, we note that the current approach can also be used when the target is best described by a density matrix, by augmenting the model with additional model wavefunctions weighted by their population factors.

\subsection{Forward mapping} \label{sec:scattering_theory}
The forward optimisation exploits that the direct problem, i.e.\ the forward mapping,
\begin{equation} \label{eq:forward-mod}
    \ket{\Psi_{\mathrm{mod}}\left(t\right)} \; \xrightarrow{M_j} \; X_\mathrm{mod}^j(t),
\end{equation}
is mathematically well-defined and has stable solutions. For each type of experimental observable the mapping $M_j$ will be different and based on different theoretical approximations and computational techniques. For instance, there is an extensive body of work on the prediction of time-resolved photoelectron spectra\cite{Worth2014,Kuhlman2012} with available techniques ranging from approximate Dyson orbitals calculations\cite{tudorovskaya_effects_2018,Krylov2007} to the highly accurate R-matrix method.\cite{UKRMOL2020} The method used to calculate the x-ray scattering and electron diffraction signals in the current paper will be discussed in Section \ref{sec:comp}.

\section{Computational Methods\label{sec:comp}}

\begin{figure}[h!]
    \captionsetup[subfigure]{justification=centering}
    \begin{subfigure}{\linewidth}
        \includegraphics[width=\linewidth, scale=1]{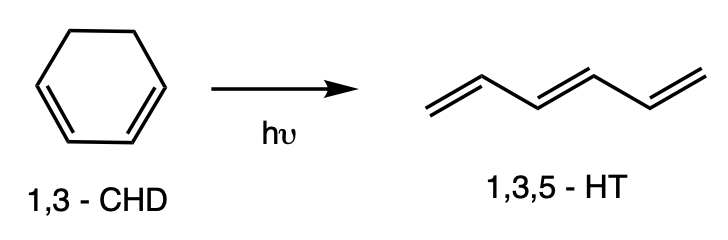}
        \caption{Ring-opening of 1,3-cyclohexadiene (CHD) yields linear HT molecules.}
        \label{fig:chd_reaction}
    \end{subfigure}
    \vskip 1ex
    \begin{subfigure}{\linewidth}
        \includegraphics[width=\linewidth, scale=1]{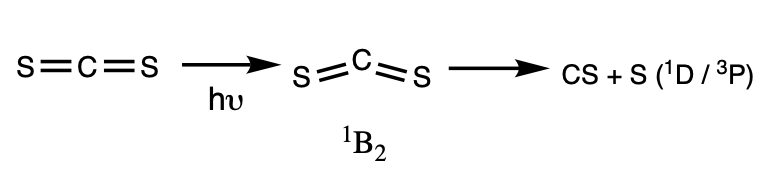}
        \caption{Dissociation of \ce{CS2} yields singlet or triplet sulphur atoms.}
        \label{fig:cs2_reaction}
    \end{subfigure}
    \caption{Schematic of the two photochemical processes probed in the experiments for which data is inverted in this paper, via a basis of trajectory basis functions (TBFs).}
    \label{fig:reaction-schematic}
\end{figure}

\subsection{Trajectory basis functions}

\subsubsection{1,3-cyclohexadiene ring-opening}
The ring-opening reaction of 1,3-cyclohexadiene (CHD) to 1,3,5-hexatriene (HT), shown schematically in Figure~\ref{fig:chd_reaction}, is a prototypical Woodward-Hoffman photoinduced electrocyclic reaction.\cite{deb_ultrafast_2011} It has been the target for a large number of pioneering time-resolved experiments that include UXS, UED, and time-resolved spectroscopies.\cite{minitti_toward_2014, minitti_imaging_2015, dudek_ultrafast_2001, cheng_9_2001, buhler_ultrafast_2011, deb_ultrafast_2011, adachi_direct_2015, pemberton_1b_2015, tudorovskaya_effects_2018, ruddock_deep_2019, wolf_photochemical_2019, filatov_structural_2020, karashima_ultrafast_2021-1} Upon absorption of a $267$ nm photon the molecule undergoes a $\pi \rightarrow \pi^{*}$ transition to the steeply sloped 1B electronic state, glancing the conical intersection with the 2A electronic state while staying on the adiabatic potential energy surface. Passage through the conical intersection to the electronic ground state returns the molecule to the original ring-closed CHD or breaks a \ce{C-C} bond to yield the ring-open HT. A distribution of various \textit{cis}$\left(Z\right)$/\textit{trans}$\left(E\right)$ HT isomers are observed given the high internal energy of the system.

Semiclassical trajectories, TBFs, for the dynamics are calculated using the {\it{ab initio}} Ehrenfest (AIMCE) method.\cite{minitti_toward_2014,minitti_imaging_2015} The electronic structure calculations use the {\it{ab initio}} package MOLPRO,\cite{MOLPRO-WIREs} which supplies the forces and the nonadiabatic couplings at 3SA-CAS(6,4)-SCF/cc-pVDZ level of theory for the ground and excited states. A set of 100 TBFs, propagated for 200~fs, with initial conditions sampled from a Wigner distribution in the Franck–Condon region of equilibrium ground state CHD provide the basis for the trajectory-fitting procedure.

\subsubsection{\ce{CS2} photodissociation}
The photodissociation of \ce{CS2}, shown schematically in Figure~\ref{fig:cs2_reaction}, has been the subject of numerous time-resolved experiments.\cite{razmus_multichannel_2022, townsend_b21u1_2006, bisgaard_time-resolved_2009, hockett_time-resolved_2011, fuji_excited-state_2011, brouard_ultraviolet_2012, spesyvtsev_observation_2015, yang_imaging_2015, horio_real-time_2017, wang_monitoring_2017, smith_mapping_2018, warne_time_2021, gabalski_transient_2022} Upon excitation to the $^1B_2$ ($^1 \Sigma_u^+$) state rapid bending and stretching vibrational motion is observed. More complex excited state dynamics ensues, exhibiting a striking competition between internal conversion (nonadiabatic couplings) and intersystem crossing (spin-orbit coupling), resulting in two dissociation channels that yield either \ce{S}($^1D$) or \ce{S}($^3P$) sulphur atoms, with the triplet product dominant.

Semiclassical trajectories (TBFs) to be used in the forward optimisation are calculated using the SHARC surface-hopping package.\cite{mai_general_2015, mai_nonadiabatic_2018} The trajectories start on the optically bright $^1\Sigma_u^+$ state with a total energy corresponding excitation by a 200 nm pulse and initial coordinates sampled from the ground state Wigner distribution. The forces and nonadiabatic couplings are calculated using SA8-CASSCF(10,8)/SVP electronic structure theory using MOLPRO\cite{MOLPRO-WIREs} and 197 trajectories are propagated for $1$~ps. Further information can be found in \textit{SI Section~\ref{sec:ued_treatment}}.

\subsection{Scattering observables}
The forward mapping required to compare the model to the UXS and UED  data is summarised below. Although scattering signals can be calculated from first principles for gas-phase molecules\cite{Techert2006,northey_ab_2014,northey_elastic_2016,moreno_carrascosa_ab_2017,Moreno2017,moreno_carrascosa_ab_2019,Parrish2019,zotev_excited_2020,Moreno2022} and recent x-ray scattering measurements have demonstrated that given sufficient accuracy different electronic states can be resolved,\cite{BenNun1996,KirranderJCP2012,stankus_ultrafast_2019,yong_observation_2020} the experimental data considered in this paper can be modelled using the significantly simpler independent atom model (IAM), which approximates the scattering as a summation over the coherent scattering from isolated atoms.

The instantaneous energy-integrated total scattering cross section into the solid angle $d\Omega$ is then,\cite{kirrander_ultrafast_2016,Minas2017,kirrander_fundamental_2017}
\begin{equation} \label{eq:cross-section}
    \frac{d\sigma}{d\Omega} = \left(\frac{d\sigma}{d\Omega}\right)_x I_\mathrm{mod}(\mathbf{q}),
\end{equation}
where $\mathbf{q}=\mathbf{k}_0-\mathbf{k}_1$ is the scattering vector expressed in terms of the wave vectors of the incoming and outgoing photon or electron.\footnote{The scattering (momentum transfer) vector $\mathbf{q}$ is denoted $\mathbf{s}$ in electron scattering.} For x-ray scattering, the prefactor $(d\sigma/d\Omega)_x$ is the differential Thomson scattering cross section $(d\sigma/d\Omega)_{\mathrm{Th}}$, here made to include the $|\mathbf{e}_0\cdot\mathbf{e}_1^*|$ polarisation factor, while for electron scattering it is the Rutherford cross section $(d\sigma/d\Omega)_{\mathrm{Rh}}$ which here includes the scaling factor $s^{-4}$.\cite{mott_scattering_1930, bethe_zur_1930} Note that the expression above does not account for the duration of the x-ray pulse, which is accounted for via the temporal convolution discussed in Section \ref{sec:apparatus-mapping}.

Using the diagonal BAT approximation and assuming time-independent expansion coefficients $\norm{\mathbf{w}}=1$, the scattering intensity $I(\mathbf{q})$ can be calculated as,
\begin{equation}
    I_{\mathrm{mod}}\left(\mathbf{q},t,\mathbf{w}\right) = \sum_{n=1}^{N_{\mathrm{TBF}}} w_n I_n\left(\mathbf{q}, \mathbf{R}_n(t)\right).
\end{equation}
More general expressions that account for the full wavefunction in eq~\ref{eq:tbf_approx}, including the non-local nature of the individual TBF nuclear wavepackets, have been derived previously.\cite{kirrander_ultrafast_2016} Continuing with the current simplified form, sufficient for our present needs, yields the scattering intensity as,
\begin{equation}
   I_n(\mathbf{q}, \mathbf{R}_n(t)) = |f(\mathbf{q},\mathbf{R}_n(t))|^2 + S_\mathrm{inel}(q), 
\end{equation}
where $S_\mathrm{inel}(q)$ is the inelastic scattering, which is independent of molecular geometry and given by an incoherent summation over the atomic contributions,   
\begin{equation}
    S_\mathrm{inel}(q) = \sum_{A=1}^{N_\mathrm{at}} S_A(q),
\end{equation}
with $N_\mathrm{at}$ the number of atoms. The corresponding elastic contribution is given by the form factor $f(\mathbf{q},\mathbf{R}_n(t))$,
\begin{equation} \label{eq:IAM}
    f(\mathbf{q},\mathbf{R}_n(t)) = \sum_{A=1}^{N_\mathrm{at}} f_A(q) e^{\imath \mathbf{q} \mathbf{R}_{nA}(t)},
\end{equation}
where $f_A(q)$ are the atomic form factors and $\mathbf{R}_{nA}(t)$ the position vector for atom $A$ in trajectory $n$. Both $f_A(q)$ and $S_A(q)$ are tabulated.\cite{IntTabCryVolC} The form factors for electron scattering are $f^\mathrm{e}=Z_A-f^\mathrm{x}$, where $Z_A$ is the atomic number and $f_A^{\mathrm{x}}$ the x-ray scattering form factor.\cite{mott_scattering_1930, bethe_zur_1930} For high energy electron scattering, notably MeV-UED, it is sometimes necessary to use form factors with relativistic corrections.\cite{salvat_elastic_1991, salvat_elsepadirac_2005} 

When the target is a gas of anisotropic molecules, rotational averaging of eq~\ref{eq:IAM} results in,\cite{debye_zerstreuung_1915}
\begin{multline}\label{eq:IAM_general}
        I\left(q, \mathbf{R}_{n}(t)\right) = \sum_{A,B}^{N_\mathrm{at}} f_{A}\left(q\right) f_{B}\left(q\right) \frac{\sin{\left(q R_{nAB}(t) \right)}}{q R_{nAB}(t)}\\
        %j_{0}\left(q R_{AB}\right) 
        \;\;\;\;+ S_{\mathrm{inel}}\left(q\right),
\end{multline}
with the distance $R_{nAB}(t) = |\mathbf{R}_{nA}(t) - \mathbf{R}_{nB}(t)|$ between atoms $A$ and $B$ in trajectory $n$.

\subsection{Apparatus mapping} \label{sec:apparatus-mapping}
In this section we discuss the {\it{apparatus mapping}} $S_j$ in eq~\ref{eq:predicted-data} that is required to match the forward mapped signal $X^j_{\mathrm{mod}}(t)$ from eq~\ref{eq:forward-mod} to the experimentally observed signal. 

%For the scattering data considered in this paper, the temporal convolution due to limited time-resolution is most important, while the experimental resolution in the momentum transfer $q$ is such that further convolution is not necessary. Furthermore, the assumed excitation fraction and the temporal alignment of theory and experiment must be included as discussed below. Finally, we present the appropriate target function expressed in terms of percentage differences of the experimental and model-predicted signal. 

\subsubsection{Temporal alignment and convolution}
In the scattering experiments, the time-zero is roughly calibrated by the instrument, with the exact time-zero inferred from the observed data. The absence of independent validation means that one must check the alignment of the temporal axes in the experiment and the model.\footnote{Absolute changes on the experimental time-axis are accurate, trivially so for the model/theory.} We define the relationship between the experimental and model time axes as $t' = t + t_0$, where $t'$ is the experiment, $t$ the model, and $t_0$ the time shift. The temporal alignment $t_0$ is one of the global parameters optimised.

A Gaussian convolution in the temporal domain is included as,
\begin{equation} \label{eq:conv}
    I_{\mathrm{mod}}^{\mathrm{conv}}(q,t) = \int_{-\infty}^{\infty} I_{\mathrm{mod}}(q,t^{\prime\prime})\;G(t-t^{\prime\prime})\;dt^{\prime\prime},
\end{equation}
where $G(t) = b_c\exp{(-a_ct^2)}$ mimics the instrument response function, with the normalisation constant given as $b_c=\sqrt{4\ln2/(\tau_c^2\pi)}$ and $\tau_c$ the full-width half-maximum (FWHM). The convolution mainly equates to the cross-correlation of the pump and the probe, effectively compensating for the $\delta(t)$-pulse excitation approximation used to generate the TBFs. In practice, the instrument response function also accounts for other limits on the temporal resolution, such as temporal jitter. 

%As a consequence of the convolution, the excitation fraction $\gamma$ and the $t_0$ determination are intertwined during the optimisation (see below). 

Finally, we note that the experimental resolution with respect to the momentum transfer $q$ is such that no convolution of the model is required. The amount of structural information in the signal is limited by the $q$-range, $q\in[q_\mathrm{min},q_\mathrm{max}]$, measured in the experiment.\cite{kirrander_fundamental_2017, minitti_imaging_2015}

\subsubsection{Percent difference signal}
The experimental signal is considered in the percent difference form to minimise systematic multiplicative errors, 
\begin{equation}\label{eq:general_pd}
    \%\Delta I_{\mathrm{exp}}\left(q, t^{\prime} \right) = 100\frac{I_{\mathrm{on}}\left(q,t^{\prime}\right) - I_{\mathrm{off}}\left(q, t^{\prime}_{j} \ll t_0 \right)}{I_{\mathrm{off}}\left(q, t^{\prime}_{j} \ll t_0 \right)},
\end{equation}
where $I_{\mathrm{on}}\left(q,t\right)$ is the optically pumped {\it{'laser-on'}} signal and $I_{\mathrm{off}}\left(q, t^{\prime}_{j} \leq t_0 \right)$ is the static {\it{'laser-off'}} reference signal measured at delay times $t'\ll t_0$. The theoretical equivalent is calculated from the model wavefunction $\ket{\Psi_{\mathrm{mod}}\left(t\right)}$ and defined as,
\begin{equation}\label{eq:general_pd_th}
    \%\Delta I_{\mathrm{mod}}^\mathrm{conv}\left(q, t \right) = 100\gamma\frac{I_{\mathrm{mod}}^{\mathrm{conv}}\left(q,t\right) - I_{\mathrm{off}}^{\mathrm{th}}\left(q\right)}{I_{\mathrm{off}}^{\mathrm{th}}\left(q\right)},
\end{equation}
where the excitation fraction $\gamma$ scales the intensity according to the implicit degree of excitation. The theoretical {\it{'laser-off'}} signal $I_{\mathrm{off}}^{\mathrm{th}}\left(q\right)$ is calculated from a suitable reference geometry, or more accurately using a Wigner distribution of the system in its ground state at the equilibrium geometry. In some cases, it may be necessary to modify the definition in eq~\ref{eq:general_pd_th} to scale the signal by the ratio of the integrated intensity of the {\it{'laser-on'}} and {\it{'laser-off'}} signal, or replace the uniform excitation fraction $\gamma$ with a $q$ dependent excitation fraction $\gamma\left(q\right)$. Such modifications are discussed in the \textit{SI}.

\subsection{Target function and confidence matrix}
The experimental percentage difference signal $\%\Delta I_{\mathrm{exp}}\left(q,t\right)$ and the signal predicted from the model $\%\Delta I_{\mathrm{mod}}\left(q,t,w\right)$ give the target function introduced in eq~\ref{eq:target_function} the following specific form,
\begin{multline} \label{eq:tfunc}
    F(\mathbf{w},\mathbf{c}) = \sum_{i,j} \left|\%\Delta I_{\mathrm{exp}}(q_i,t_j^\prime)\right. \\
\;\;\;\;\;\;\;\;\;\;\;\; \left. - \%\Delta I_{\mathrm{mod}}(q_i,t_j,\mathbf{w})\right|^{2} p_\mathrm{conf}(q_i,t_j^\prime),
 \end{multline}
where $\mathbf{w}=(w_1,w_2,\ldots w_{N_{\mathrm{TBF}}})$ are the normalised trajectory weights $\|\mathbf{w}\|=1$, and $\mathbf{c}$ consists of additional global parameters to be optimised, such as e.g.\ the excitation fraction $\gamma$ and the time-shift $t_0$. In principle, more complicated global parameters designed to off-set shortcomings in the quality of the TBFs could be included, such as time-warping to offset inaccuracies in the kinetic energy. The double sum runs over all experimental data points, identified by $q_i$ momentum transfer and the temporal coordinates $t_{j}^{\prime}=t_{j}+t_0$. 

To avoid overfitting of inherently noisy experimental data, the target function includes a confidence matrix. The matrix $p_\mathrm{conf}(q_i,t_j^\prime)$ weights data points, identified by their value of the momentum transfer $q_i$ and time bin $t_j^{\prime}$, according to the experimental confidence in the accuracy of that data point. For instance, data points subject to poor statistics or systematic errors are given smaller weight while points that are known or expected to be accurate are weighted accordingly. The exact form of $p_\mathrm{conf}(q_i,t_j^\prime)$ depends on the data set, as discussed in the \textit{SI}.  

If needed, one can define a confidence threshold $p_{\mathrm{conf}}^{\mathrm{min}}$, such that all points $p_{\mathrm{conf}}\left(q_i, t_j^{\prime}\right) \leq p_{\mathrm{conf}}^{\mathrm{min}}$ are zero, thus excluding them from the optimisation. The advantage of excluding poor quality points from the optimisation must be balanced against the reduced size of the experimental data set.

\subsection{Global optimisation}
In the ideal scenario, the experimental signal has sufficient quality (high temporal resolution, large $q$-range, and excellent signal-to-noise) that simultaneous optimisation of all model parameters is possible. Since the number of global parameters is small, the most straightforward approach is to optimise the trajectory weights $\mathbf{w}$ for fixed global parameters $\mathbf{c}$, and to determine from a scan of reasonable values of $\mathbf{c}$ the best overall solution.

For each particular $\mathbf{c}$, a pool of $N_\mathrm{init}$ initial weights $\{\mathbf{w}\}_\mathrm{init}$ are Monte Carlo sampled with the values of the individual weights in each initial set $w_i\in[0, 1]$. The target function given in eq~\ref{eq:tfunc} is then minimised utilising a non-linear trust-region reflective least-squares algorithm for each initial set. Among the optimisations that converge to a local minimum, within the tolerance constraints, the best minimum is considered a candidate for the global minimum. The number of initial conditions $N_\mathrm{init}$ generated in the weight space is increased until apparent convergence is achieved. The convergence of the optimisation with respect to $N_{\mathrm{init}}$ is discussed in \textit{SI}, Section~\ref{sec:condition_convergence}. In challenging cases, an iterative approach using more targeted sampling can be used, which is discussed in the \textit{SI}.

\subsection{Two-step optimisation}\label{sec:two_step}
When the experimental data is of lower quality, more stable solutions are found via a two-step optimisation procedure. The first step ensures that global parameters, such as $t_0$ and $\tau_c$, are determined as accurately as possible before, in the second step, optimisation of the trajectory weights is attempted.

\subsubsection{Step 1: Global parameters}\label{sec:t0_fit}
First, we identify the strongest features in the data by inspection of the confidence matrix $p_\mathrm{conf}(q_i,t_j^\prime)$. This allows us to minimise the negative effects of noise on the optimisation. Selecting the strongest feature, we define the net integrated percentage difference signal as,
\begin{equation}\label{eq:integrated_signal}
    \%\Delta I^{\mathrm{int}}\left(t\right) = \int_{q^{\mathrm{conf}}_{\mathrm{min}}}^{q^{\mathrm{conf}}_{\mathrm{max}}} \% \Delta I\left(q, t\right) dq,
\end{equation}
where $q^{\mathrm{conf}}_{\mathrm{min}}$ and $q^{\mathrm{conf}}_{\mathrm{max}}$ are the bounds on the section of highest confidence. The global parameters $\mathbf{c}$ are then optimised against this integrated signal, $\%\Delta I^{\mathrm{int}}_{\mathrm{exp}}\left(t\right)$. Since this step must be performed independently from the optimisation of the trajectory weights $\mathbf{w}$, we assume that the $t\approx t_0$ wavefunction can be approximated by an equally weighted sum of TBFs. This is reasonable given the limited dispersion and dephasing in the wavepacket at early times. 

The $t_0$ is inherently linked to $\tau_c$ and $\gamma$ since the pulse width $\tau_c$ affects the onset of the signal and $\gamma$ scales the strength of the signal. Therefore, for different combinations of $t_0$ and $\tau_c$, the sum of square error between the experimental and model integrated difference signal in eq~\ref{eq:integrated_signal} is minimised with the $\gamma$ excitation fraction as the free parameter.

\subsubsection{Step 2: TBF weights\label{sec:weight_opt}}
The second step in the optimisation takes the best values of $\left(t_0, \tau_c, \gamma \right)$ from the previous step with the goal of identifying the optimal trajectory weights $\mathbf{w}$. Since the $\gamma$ in step one is determined on the assumption of equally weighted TBFs, the parameter $\gamma$ is reoptimised alongside the weights $\mathbf{w}$ in this second step. If the final value of $\gamma$ diverges significantly from its initial value, this may indicate that the parameters determined in the first step are not optimal. In order to ensure self-consistency, it is advisable to repeat the second step optimisation for several different sets of global parameter values $\left(t_0, \tau_c, \gamma \right)$ which correspond to good fits in the first step. The procedure is complete once the best sets of these values from step one and two agree, and when the target function $F(\mathbf{w},\mathbf{c})$ has converged.

\subsection{Metrics of fit quality}\label{sec:stats}
In addition to the target function $F(\mathbf{w},\mathbf{c})$, it is useful to have other measures of the quality of the fit. The improvement in overall agreement between the unweighted and weighted model function can be quantified by the relative absolute error (RAE),
\begin{equation} \label{eq:rae}
\begin{split}
    &\mathrm{RAE} =  \\ & \frac{1}{N_t}\sum_{t}\frac{|\frac{1}{N_q} \sum_{q} \%\Delta I_{\mathrm{mod}}(q, t) - \%\Delta I_{\mathrm{exp}}(q,t)|} {|\frac{1}{N_q}\sum_{q} \%\Delta I_{\mathrm{th}}(q, t) - \%\Delta I_{\mathrm{exp}}(q,t)|},
\end{split}
\end{equation}
where $N_t$ is the number of time steps, $N_{q}$ the number of points in $q$, and $\%\Delta I_{\mathrm{th}}(q, t)$ the theoretical signal calculated from the theoretical wavefunction not subjected to any bias or optimisation. The RAE measure, as defined above, is independent of the size of the $N_t\cross N_q$ grid. Another helpful metric is the root mean squared error (RMSE) defined as,
%\mathrm{RMSE} = \frac{1}{N_qN_t} \sum_{q,t} \sqrt{\left(\%\Delta I_{\mathrm{mod}}(q, t) - \%\Delta I_{\mathrm{exp}}(q,t)\right)^{2}}.
\begin{equation}\label{eq:rmse}
    \mathrm{RMSE} = \sum_{q,t} \frac{|\%\Delta I_{\mathrm{mod}}(q, t) - \%\Delta I_{\mathrm{exp}}(q,t)|}{N_qN_t}.
\end{equation}

Particular solutions can also be characterised using the variance of the $N_{\mathrm{TBF}}$ weights from a single optimisation, {\it{i.e.}}\ the spread of weights resulting from one of the $N_{\mathrm{init}}$ initial conditions (typically the initial condition that yields the best solution for a specific set of global parameters), defined as,
\begin{equation}\label{eq:weight_var}
    v_w^{2} = \sum_{n=1}^{N_{\mathrm{TBF}}}| w_n - \left<\mathbf{w}\right> |^{2},
\end{equation}
where $\left<\mathbf{w}\right>$ is the mean of the weights. In addition, to compare different solutions we define the distance $D_b$ from the overall best set of optimised weights,
\begin{equation}\label{eq:opt_dist}
    D_{b}^{2} = |\mathbf{w}_{b} - \mathbf{w}_{\mathrm{best}}|^{2},
\end{equation}
which describes the distance for a particular solution $b$ from the best set of optimised weights.

\section{\label{sec:methods}Experimental data}

\subsection{Ultrafast x-ray scattering}
\begin{figure}[h!]
    \captionsetup[subfigure]{justification=centering}
        \includegraphics[width=\linewidth]{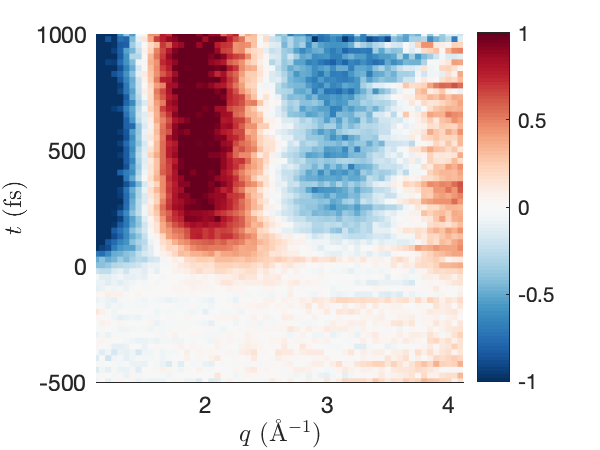}
        \centering
        \caption{A subsection of the experimental UXS signal $\%\Delta I_\mathrm{exp}(q,t)$ for CHD from ref~\citenum{minitti_imaging_2015}. The experimental signal has been shifted in time so that it is centered at $t=0$, instead of the $t' = -110$~fs along the original raw experimental time axis.}
        \label{fig:dIexp_chd}
\end{figure}

The ultrafast x-ray scattering (UXS) data for the ring-opening reaction of CHD shown in Figure~\ref{fig:dIexp_chd} is taken from ref~\citenum{minitti_imaging_2015} (further details in the \textit{SI}, \textit{Section~\ref{sec:uxs_treatment}}). The confidence matrix $p_\mathrm{conf}$ in eq~\ref{eq:target_function} in this case is based on the number of photon hits per frame (\textit{SI}, eq~\ref{eqn:conf_mat_xray}). Due to a long interaction region in these experiments, we include a $q$-dependent excitation fraction $\gamma\left(q\right)$ which is indirectly optimised by allowing it to be uniformly scaled by a factor $x$ such that it yields the scaled excitation fraction $\gamma_{x}\left(q\right)$ (for more details see the \textit{SI}). Accounting for the convolution of the signal as in eq~\ref{eq:conv} and the subsequent temporal binning into bins of size $\Delta t = 25$~fs, the total length of the signal used becomes 275~fs.

The quality of the experimental data allows a one-step global optimisation in which we scan over a range of $t_0$ values and $x$ scaling factors, where $t_0 \in \left[-38, -14\right]$ and $x \in \left[0.7, 1.3\right]$.
The duration of the pump and probe pulse were measured from experiment as 60~fs and 30~fs respectively, which fixes the value of $\tau_c$.

\subsection{Ultrafast electron diffraction}
The ultrafast electron diffraction (UED) data for the photodissociation of \ce{CS2} is taken from ref~\citenum{razmus_multichannel_2022} (further details in the \textit{SI}, \textit{Section~\ref{sec:ued_treatment}}). The data does not support the real-space pair distribution function (PDF) analysis obtained by a sine-transform of the modified scattering signal $\Delta sM$ that is common in UED (see \textit{SI}, \textit{Section~\ref{sec:pdf_limitations}}). Instead, the signal is evaluated in the percent difference form as per eq~\ref{eq:general_pd}. 

\begin{figure}[h!]
    \captionsetup[subfigure]{justification=centering}
        \includegraphics[width=\linewidth]{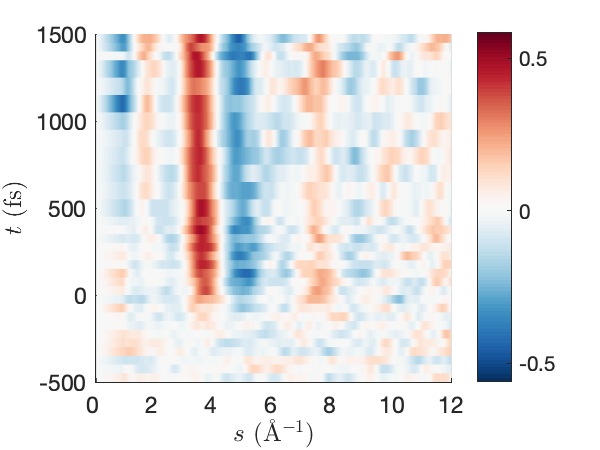}
        \centering
        \caption{A subsection of the UED signal $\%\Delta I_\mathrm{exp}(s,t)$ for \ce{CS2} from ref~\citenum{razmus_multichannel_2022}. The signal is initially centred on $t' = -120$~fs.}
        \label{fig:dIexp_cs2}
\end{figure}

The experimental signal $\%\Delta I_\mathrm{exp}(s,t)$ in Figure~\ref{fig:dIexp_cs2} displays a strong enhancement band (red) next to a strong depletion band (blue) in the range $3.5 < s <6.0$~\AA$^{-1}$. Also note less intense enhancement/depletion features for $s<2$~\AA$^{-1}$ which appear at later times and which correlate with the onset of strong dissociation. 

Given the noise and limited temporal resolution in the data, we carry out a two-step optimisation. In the first step, we fit $t_0$ using eq~\ref{eq:integrated_signal} with bounds $[s_{\mathrm{min}}, s_{\mathrm{max}}] = [2.8, 4.2]$ \AA$^{-1}$. This procedure is repeated in the range $t_0 \in \left[-16, 83\right]$~fs and $\tau_c \in \left[150, 250 \right]$~fs. The best global parameters are then used in the determination of the weights $\mathbf{w}$. Initial conditions for $\mathbf{w}$ are generated both using unbiased Monte Carlo sampling and using the biased iterative sampling procedure described in \textit{SI}, \textit{Section.~\ref{sec:weights}}. In either case, the confidence matrix $p_\mathrm{conf}$ is based on estimated experimental standard deviations (see \textit{SI}, eq~\ref{eq:conf_mat_ued}).

\section{Results and Discussion}\label{sec:results}

\subsection{CHD ring-opening (UXS)}

\begin{figure}[hbt!]
    \centering
    \includegraphics[width=\linewidth]{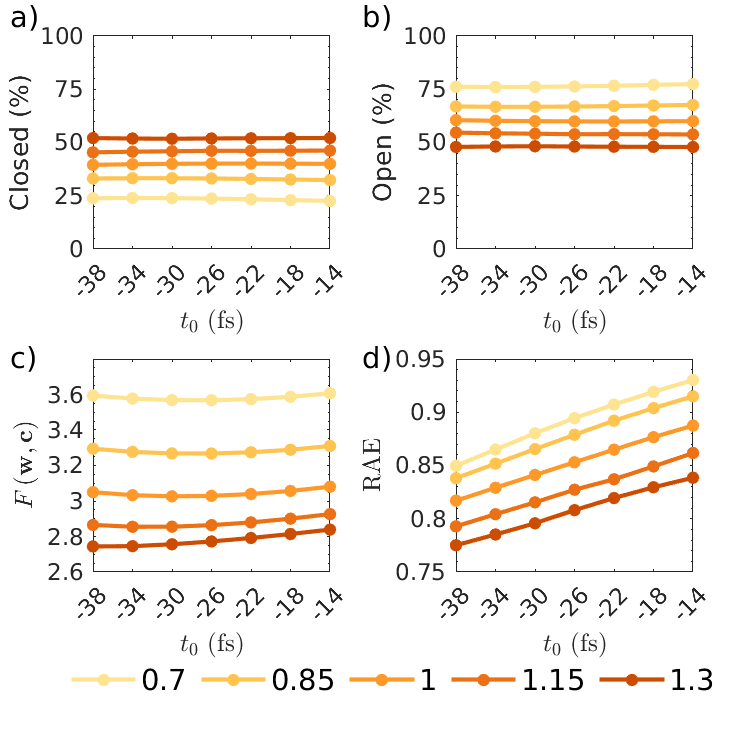}
    \caption{Convergence in the CHD optimisation, showing the a) ring-closed fraction, b) ring-open fraction, c) the target function and d) the relative absolute error (RAE) as a function of $t_0$. Each colour represents a different value of the scaling factor $x$, ranging from 0.7 to 1.3.}
    \label{fig:chd_global_conv}
\end{figure}

For the CHD reaction, the data is of sufficient quality that a global one-step optimisation is feasible. Optimisation yields the best fit parameters $\left[t_0, x\right] = [-38\;\mathrm{fs}, 1.3]$, with a RAE of $0.775$. Recall, the scaling factor $x$ is optimised as to uniformly scale the $q$ dependent excitation fraction such that, $\gamma_{x}\left(q\right) = \gamma\left(q\right) x$. The values of $\gamma\left(q\right)$ and the resulting optimised $\gamma_{x}\left(q\right)$ over the available $q$ range, can be seen in \textit{SI} \textit{Figure~\ref{fig:chd_xfrac}}. The convergence with respect to $x$ and the $t_0$ shift in Figure~\ref{fig:chd_global_conv} shows that more negative $t_0$ shifts and larger $x$ result in lower values of the target function $F\left( \mathbf{w}, \mathbf{c}\right)$ and better RAE's (with the trend more pronounced for higher $x$). Generally, $F\left( \mathbf{w}, \mathbf{c}\right)$ is more sensitive to $x$ than $t_0$.

The change in the fraction of ring-open trajectories is small as $t_0$ changes, which is unsurprising given that the majority of the ring-opening occurs in a concerted fashion within the first 140~fs.\cite{deb_ultrafast_2011} Current literature values are found in the range 40-60\%.\cite{deb_ultrafast_2011} We note that higher values of $x$ redistribute some of the ring-open weights to ring-closed weights. Since one does not physically expect the fraction of open to closed trajectories to change as a function of the excitation fraction, this is an artefact which stems from the fact that the percent difference signal is stronger for ring-open molecules than ring-closed, due to the reference equilibrium CHD structure being ring-closed. In principle, this could be addressed by a different choice of reference structure.

\begin{figure}[h!]
    \centering
    \includegraphics[width=\linewidth]{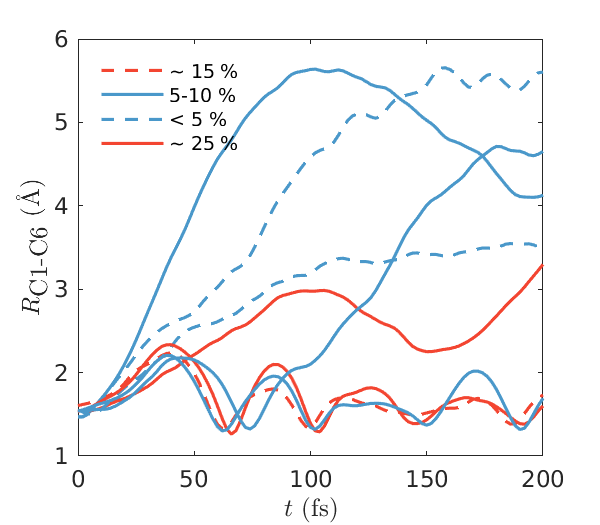}
    \caption{\ce{C1-C6} distances for the eight CHD trajectories with a weight greater than 1\%. There are three distinct classes of trajectories: ring-closed, direct ring-opening and indirect (delayed) ring-opening.}
    \label{fig:chd_distances}
\end{figure}

The eight dominant trajectories in the final solution are summarised in Table~\ref{tab:chd_tab}. These account for almost all the weights. The trajectories can be visualised in terms of their characteristic \ce{C1-C6} bond distance as shown in Figure~\ref{fig:chd_distances}. Three main families of trajectories are observed: direct ring-opening, a slower indirect ring-opening which undergoes several \ce{C1-C6} stretches before breaking the bond, and finally ring-closed paths with initially strong oscillations in the \ce{C1-C6} bond which are damped out as the energy disperses across all motions. In total, the ring-opening and closed trajectories have a weight of 52~\% and 48~\% respectively. 

The ground state HT has several \textit{cis-/trans-} $\left(Z\right)/\left(E\right)$ isomers. Out of the 48\% HT product, we predominately observe the presence of cEc-HT (8.7\%) and cZc-HT (34.5\%) isomers. Due to the length of the 275~fs temporal window used in the fit it is no surprise we can not clearly detect the tZt-HT isomer. However, we note that we observe a small fraction (3.4\%) with a configuration between cZt and tZt. With data available over a longer temporal range, one could better refine the ground state dynamics and eventually observe as the system settles into a thermal equilibrium between different HT isomers.

\begin{table}[h!]
    \centering
    \begin{tabular}{c|c}
         Weight (\%) & Type  \\
         \hline
         28.4 & open (indirect) \\
         27.3 & closed \\
         15.1 & closed \\
         $\;\;\;$9.70 & closed \\
         $\;\;\;$8.70 & open \\
         $\;\;\;$6.05 & open (indirect) \\
         $\;\;\;$3.38 & open \\
         $\;\;\;$1.35 & open \\
         
    \end{tabular}
    \caption{Weights of the dominant trajectories from the forward optimisation for CHD, along with their type.}
    \label{tab:chd_tab}
\end{table}

\begin{figure}[hbt!]
    \centering
    \includegraphics[width=\linewidth]{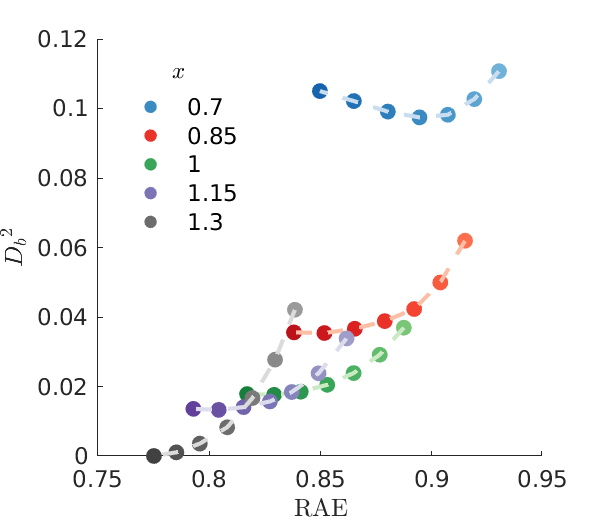}
    \caption{Square distances $D_b^2$ (eq~\ref{eq:opt_dist}) of the weights from a series of different optimisations relative the best weights with $\left[t_0, x \right] = \left[-38, 1.3 \right]$, plotted against the relative absolute error (RAE). Each shade represents a different $t_0$ ranging from -38 to -14~fs (the darker the shade, the earlier the $t_0$ shift).}
    \label{fig:chd_best_var}
\end{figure}

In Figure~\ref{fig:chd_best_var}, one can see there is little variation from the {\it{best}} set of weights if one examines solutions with more negative $t_0$ shifts or larger $x$. The cluster of very good optimisations are all found within a small radius of $D_{b}^2$ from the optimal solution, and have RAE's in the range 0.77-0.82 with only a slight re-weighting of similar trajectories. The solutions are further removed when when $x = \left[0.7, 0.85\right]$, which is a consequence of the optimisation redistributing weight between ring-closed trajectories and selecting a different set of ring-opening trajectories. In addition, we note that the better optimisations (as measured by their RAE) tend to exhibit a larger variance in their set of weights, this can be seen in \textit{SI} \textit{Figure~\ref{fig:chd_var_weights}}.

%The optimisation repeatedly selects trajectories from the same pool of 11 total trajectories. Out of the eight dominant trajectories, four are selected in every optimisation, two are selected in all but two (out of 35), and the final two are selected in 14 and four out of the total 35. 

\subsection{\ce{CS2} photodissociation (UED)}
\begin{figure}[h!]
    \centering
    \includegraphics[width=\linewidth]{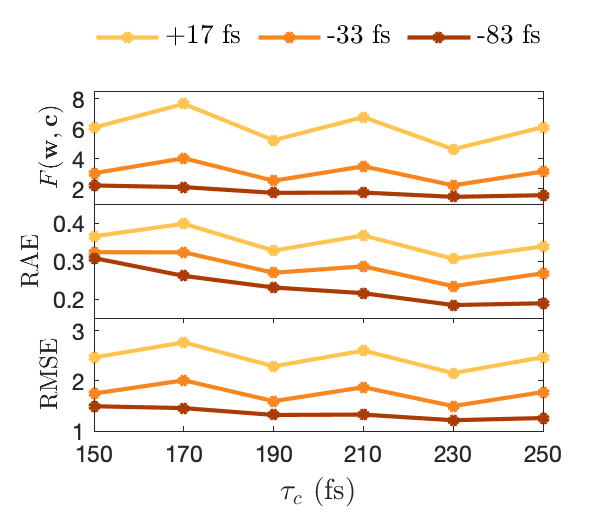}
    \caption{Convergence of the step one optimisation with respect to $t_0$ and $\tau_c$. The panels show the convergence of the target function $F(\mathbf{w},\mathbf{c})$ (top), the RAE (middle), and RMSE (bottom).}
    \label{fig:T0_conv_cs2}
\end{figure}

\begin{figure}[h!]
    \centering
    \includegraphics[width=\linewidth]{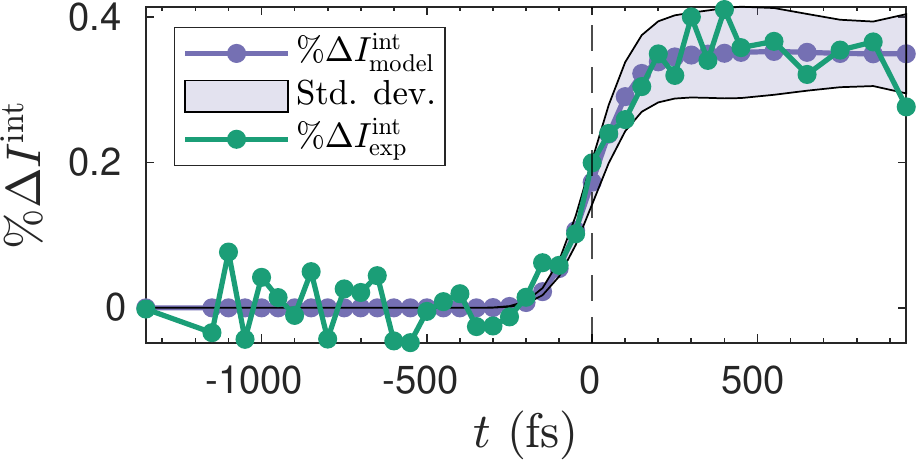}
    \caption{Final result for the step one optimisation against the integrated intensity $\%\Delta I_{\mathrm{model}}^{\mathrm{int}}$ (eq~\ref{eq:integrated_signal} with equal weights). The model is convoluted with a Gaussian with $\tau_c=230$~fs (FWHM) and the optimised value of excitation fraction is $\gamma = 3\%$. The shaded area indicates the standard deviation of the ensemble of trajectories from the model calculated assuming equal weights.}
    \label{fig:t0fit}
\end{figure}

We now turn to the more difficult case of \ce{CS2}, where the resolution forces us to employ the two-step optimisation procedure.
The first step of optimisation yields the optimal global parameters $\left[t_0, \tau_c, \gamma \right] = \left[-83\textrm{~fs}, 230\textrm{~fs}, 3\%\right]$. Figure~\ref{fig:T0_conv_cs2} shows that not only does $t_0 = -83$~fs and $\tau_c = 230$~fs result in the lowest value of $F\left(\mathbf{w}, \mathbf{c} \right)$, but also that the two complementary error measures RAE and RMSE confirm this. These parameters also yield smoother convergence across all values of $\tau_c$ in comparison to other values of $t_0 = \left[+17, -33\right]$. The resulting final best fit in step one is shown in Figure~\ref{fig:t0fit}. The model tracks the experimental data closely, especially in the important region around $t=0$, while at other times the rather scattered experimental data is contained with one standard deviation of the model, with the standard deviation calculated on the whole ensemble of trajectories relative the equally-weighted 'average' model. Also note the significant noise floor in the experimental data evident for $t\ll0$. This emphasises the importance of a robust method for fitting, inverting, and interpreting the experimental data.

For the second step of the optimisation, we take the global parameters determined in step one as the starting point for the unconstrained determination of the TBF weights $\mathbf{w}$. We also repeat the procedure for three different values of $t_0$ to check that the optimisations are consistent. During optimisation, the pre-determined value of the global parameter $\gamma$ is allowed to readjust. The value should change little but if it does change significantly this may indicate that the values from step one are suboptimal. 

\begin{figure}[h!]
    \centering
    \includegraphics[width=\linewidth]{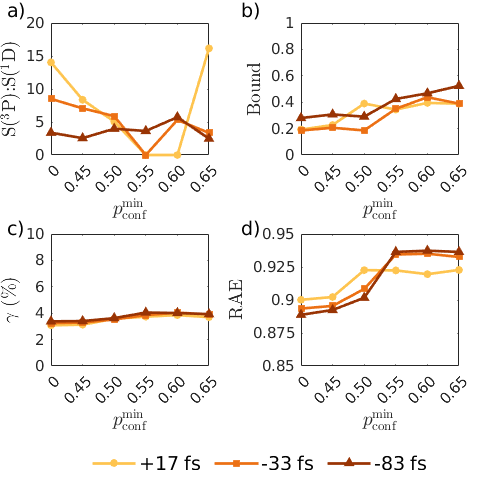}
    \caption{Convergence in \ce{CS2} optimisation (step two) as a function of confidence threshold $p_\mathrm{conf}^\mathrm{min}$ for three values of $t_0$, showing a) branching ratio, b) bound fraction at 1 ps, c) excitation fraction $\gamma$, and d) relative absolute error (RAE).}
    \label{fig:cs2_global_conv}
\end{figure}

The weighting of the diffraction signal with the confidence matrix, $p_{\mathrm{conf}}\left(s_i,t_j^\prime\right)$, is crucial to achieve a sensible fit, with significantly poorer results obtained without it. To investigate this effect, the optimisation was repeated for several values of the confidence matrix cut-off threshold, $p_\mathrm{conf}^\mathrm{min} \in [0, 0.45, 0.50, 0.55, 0.60, 0.65]$. As seen in \textit{SI Figure~\ref{fig:cs2_confmats}}, increasing the threshold corresponds to hard filtering of the data and lower thresholds correspond to including more data. For example, $p_\mathrm{conf}^\mathrm{min}=0$ includes all data, while $p_\mathrm{conf}^\mathrm{min}=0.65$ only includes the main feature in Figure~\ref{fig:dIexp_cs2}. Examining the RAE in Figure~\ref{fig:cs2_global_conv}d, it is evident that at all values of $t_0$, high $p_{\mathrm{conf}}^{\mathrm{min}}$ is detrimental to the optimisation and $p_{\mathrm{conf}}^{\mathrm{min}} = 0$ yields the best results. This suggests that including all data is beneficial as long as the data is given appropriate confidence weights. This is further confirmed by examining the physical observables resulting from the inversion, as discussed next.

Figures~\ref{fig:cs2_global_conv}a-c show the convergence of three physical observables as a function of the convergence threshold $p_{\mathrm{conf}}^{\mathrm{min}}$ for $t_0=[17,-33,-83]$ fs. The observables are not explicitly part of the fit, but are indirectly a function of the optimised weights. The branching ratio between singlet and triplet dissociation product varies significantly over the range of times and confidence thresholds, and convergence to the spectroscopy-derived literature value ($\approx3$) is only obtained for $t_0 = 83$ fs and then only at low values of confidence threshold $p_{\mathrm{conf}}^{\mathrm{min}} \rightarrow 0$. In contrast, the convergence of the fraction of bound trajectories at 1 ps is more uniform across the different $t_0$ values. Values in the range of 20-30\%, congruent with spectroscopic results, are observed for the lower confidence thresholds. A similar trend can be observed for the excitation fraction $\gamma$, which is comparatively stable across all values of $t_0$ and $p_{\mathrm{conf}}^{\mathrm{min}}$. This indicates that a suitable choice of $\tau_c$ was made and is also reflective of the fact that the value of $\gamma$ is most governed by the strongest feature in the experimental data (which is also recorded with the lowest noise).   

\begin{table}[h!]
    \centering
    \begin{tabular}{c|c}
         Weight (\%) & Type  \\
         \hline
         44.7 & Triplet \\
         17.8 & Bound \\
         10.4 & Singlet \\
         10.2 & Bound \\
         $\;\;\;\;$7.42 & Triplet \\
         $\;\;\;\;$4.42 & Singlet \\
         $\;\;\;\;$3.39 & Triplet \\
         $\;\;\;\;$1.67 & Singlet \\
         
    \end{tabular}
    \caption{Weights of the dominant trajectories in the best forward optimisation for \ce{CS2}.}
    \label{tab:cs2_tab}
\end{table}

Overall, the best optimisation yields the final values $\left[t_0, \tau_c, \gamma, p_{\mathrm{conf}}^{\mathrm{min}} \right] = \left[-83\textrm{~fs}, 230\textrm{~fs}, 3.4\%, 0 \right]$, with the convergence of $F\left(\mathbf{w}, \mathbf{c}\right)$ shown in Figure~\ref{fig:cs2_tfunc} in the \textit{SI} for reference. The initial guess of $\gamma = 3.0 \%$ from step one is reoptimised to $3.4 \%$ in the second step, with the comparatively small change suggestive of a stable solution. Examination of the optimised weights show that eight dominant trajectories account for $99.9$\% of the total weight. Two of these are bound trajectories that make up a total of $28$\%, three correspond to singlet dissociation, and two to triplet dissociation, which gives a branching ratio of singlet to triplet of 1:3.38. A summary of the eight trajectories is given in Table~\ref{tab:cs2_tab}. The contributions from each of dissociative and bound classes of trajectories to the model signal $\%\Delta I_{\mathrm{mod}}\left(s, t\right)$ can be seen in \textit{SI Figure~\ref{fig:cs2_signal_decomp}}. From this we attribute the onset of the peak below 2~\AA$^{-1}$ in the experimental signal $\%\Delta I_{\mathrm{exp}}\left(s, t\right)$ to dissociation.

\begin{figure}[h!]
    \centering
    \includegraphics[width=\linewidth]{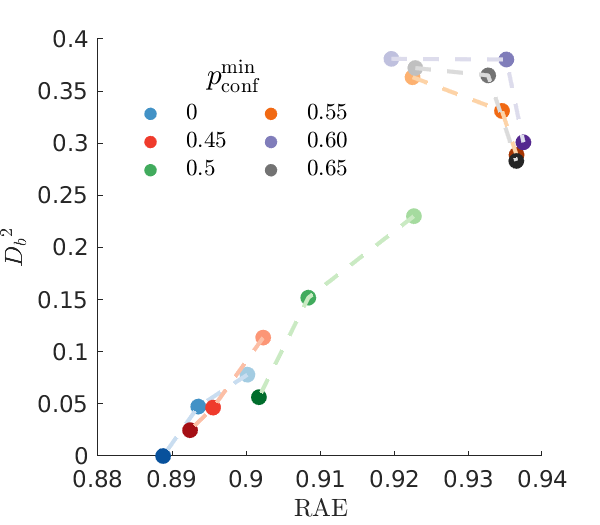}
    \caption{Squared distances $D_b^2$ (eq~\ref{eq:opt_dist}) of the final TBF weights for \ce{CS2}, examining a series of different optimisations relative the best solution with $\left(t_0, \tau_c, \gamma, p_{\mathrm{conf}}^{\mathrm{min}} \right) = \left(-83, 230, 3.4, 0 \right)$. The results are plotted with the relative absolute error (RAE) on the $x$-axis. For each value of $p_{\mathrm{conf}}^{\mathrm{min}}$, the varying opacity represents a different $t_0$ shift, ranging from -83, -33 to +17~fs from dark to light respectively.}
    \label{fig:cs2_weight_var}
\end{figure}

In Figure~\ref{fig:cs2_weight_var} we see that the set of very good solutions, as judged by their RAE value, cluster around the {\it{best}} solution as measured by the distance $D_b$ (eq\ \ref{eq:opt_dist}). Broadly, this indicates that the optimisation procedure succeeds in locating robust global optima. The clustering trend is especially evident when $p_{\mathrm{conf}}^{\mathrm{min}} \leq 0.5$ and $t_0 = \left[-83, -33\right]$. As the value of $p_{\mathrm{conf}}^{\mathrm{min}}$ increases and $t_0$ is shifted towards earlier times, the distances increase. These deviations at high values of $p_{\mathrm{conf}}^{\mathrm{min}}$ are explained by the fact that the bound population at 1~ps is over-estimated due to the omission of a distinct feature just below 2~\AA$^{-1}$ with a delay in the onset of the characteristic of dissociation (see Figure~\ref{fig:dIexp_cs2}). As with CHD, we see that the better optimisations (as measured by their RAE) also exhibit a larger variance in their set of weights, this can be seen in \textit{SI Figure~\ref{fig:cs2_var_weights}}.

%It is worth noting that generally there is little variance from the best set of weights that result from $\left(t_0, \tau_c, \gamma, p_{\mathrm{conf}^{\mathrm{0}}} \right) = \left(83, 230, 3.4, 0 \right)$. As seen in Figure~\ref{fig:cs2_weight_var} this is especially true for the lower values of confidence threshold (i.e. $p_{\mathrm{conf}}^{\mathrm{min}} \leq 0.5$, where one can also see that the variation from the best set of weights increases with earlier shifts in $t_0$. Across all of these values, the optimisation converges on the same pool of 16 trajectories. The deviation of the higher values of $p_{\mathrm{conf}}^{\mathrm{min}}$ is explained by the fact that the bound population at 1~ps is over estimated due to the omission of the feature just below 2~\AA$^{-1}$ with a delay in onset characteristic of dissociation. The deviation of the branching ratio from a reasonable range for the earlier values of $t_0$ can be understood as the mismatch in the rise of the dissociative marker below 2~\AA$^{-1}$, hence earlier $t_0$ shifts underestimate the contribution from the faster singlet dissociation channel as it shifted to later times.

It is worth noting that for the best $t_0$ and for each confidence threshold, the optimisation converges to some combination of the same subset of 16 trajectories. While only 8 of these are significant in the best optimisation where $(p_\mathrm{conf}^\mathrm{min}) = 0$, generally the weights of these trajectories show little variation as a function of the confidence threshold. Some are re-weighted significantly, for example the dominant triplet trajectory with a weight of nearly 45\% only appears in the set of trajectories once the confidence threshold is lowered and more data included. The same is true of a singlet trajectory which sees its weight increase $0\%\rightarrow4.5$\% as the confidence threshold is decreased. The opposite effect is seen for a triplet trajectory and a bound trajectory. As more data is included, these trajectories see their weight decrease $20\%\rightarrow7.5$\% and $32\%\rightarrow18$\%, respectively. Thus the contribution of some trajectories can be over-estimated and the contribution of others under-estimated when the optimisation is based on high confidence thresholds, i.e.\ smaller data sets. 

A closer examination of the distribution of trajectory weights as a function of $t_0$ shows that the relative distribution between singlet and triplet trajectories is sensitive to $t_0$. This is not surprising given that singlet and triplet dissociation is separated temporally, with the singlet dissociation appearing earlier. Shifts in the temporal alignment of experiment and model thus forces the optimisation to try to compensate by changing the relative composition of the singlet vs triplet dissociation. 

%A deeper dive into the distribution of the trajectories over all $t_0$ shifts when $(p_\mathrm{conf}^\mathrm{min}) = 0$ reveals that trajectories 14, 36, 93, 96 and 176 are picked out by all three of these optimisations. However, the inclusion of singlet trajectories 43 and 90 is unique to the best optimisation where $t_0 = 83$ fs. The inclusion of these two singlet trajectories along with the re-weighting of the triplet trajectory 36 to 45\% results in the convergence of the branching ratio to sensible literature values, as seen in Figure~\ref{fig:cs2_global_conv}. Moreover, the slight redistribution of weight amongst the bound trajectories once $t_0 = 83$ fs leads to a slight increase in the bound fraction vs $t_0 = -17$ or $+33$ fs. The filtering of different trajectories and the redistribution of these weights as $t_0$ changes and the resulting convergence to the best fit can be understood simply as the misalignment of the structure in the theoretical and experimental signals that are characteristic of dissociation.

Finally, using the iterative weight sampling procedure outlined in the \textit{SI}, we can confirm that convergence to the global minimum is achieved for the parameters $[t_0, \tau_c, \gamma, p_\mathrm{conf}^\mathrm{min}] = \left[-83~\mathrm{fs}, 230~\mathrm{fs}, 3.4\%, 0\right]$ (for further details, see the \textit{SI}).

\section{Conclusions}
In this paper, we have presented a forward optimisation method for the inversion of time-resolved data and evaluated its performance on data from actual experiments. The method matches experimental data to a model molecular wavefunction by optimising the weights of trajectory basis functions (TBFs). The TBFs provide appropriate constraints on molecular systems far from equilibrium and ensure that continuity relations are fulfilled. The method can be applied to any type of experiment for which observables can be calculated from the model wavefunction and to any molecular system for which a basis of trajectories can be generated. 

Notably, the TBFs ensure a physically sensible solution even when the inversion is underdetermined by the available data, which is the common situation in ultrafast experiments where the observable often only illuminates one specific aspect of the complex dynamics, or when the data has limited temporal, spatial, or energy resolution (for instance a limited $q$-range in scattering experiments). Importantly, we demonstrate that our approach is robust for noisy data and show that using a confidence matrix for the experimental data is far superior to excluding noisy data from the inversion. For very noisy data, we use a two-step optimisation which first optimises global parameters and then the model wavefunction. 

We apply the method to ultrafast scattering data for the molecules \ce{CS2} and CHD. Good agreement with the experimental data is found, and the model performs well also on additional quality measures introduced to evaluate the model wavefunction and the stability of the solution. The appraisal of the solution includes comparison to spectroscopic data not included for in the inversion, with the optimised model reproducing key physical properties such as the branching ratio of ring-open to ring-closed product molecules for CHD and the ratio of singlet to triplet dissociation products in \ce{CS2}.  

The procedure allows key dynamic motifs that contribute to the signal to be identified. For CHD, the method disentangles three pathways involved in the ring-opening and for \ce{CS2} it identifies a key marker of dissociation in the signal. 

Ultrafast experiments tend to include theory and simulations as part of their analysis. The current approach fits naturally into this workflow by allowing the discrepancies between theory and experiment to be identified while providing a robust interpretation of the experimental data. The forward optimisation analysis can be performed at small additional cost and provides a bridge between the theory and the experiment, and introduces a systematic strategy for addressing potential shortcomings in simulations. 

Looking ahead, forward optimisation should be applied to other observables and, more importantly, to combined data sets (e.g.\ photoelectron spectroscopy {\it{and}} scattering data). For scattering experiments specifically, higher quality data will potentially allow for the inversion to account for state-specific scattering,\cite{northey_ab_2014,moreno_carrascosa_ab_2019} inelastic effects,\cite{yang_simultaneous_2020,Zotev2020} possibly coherent mixed scattering,\cite{simmermacher_theory_2019} and, finally, alignment effects.\cite{CenturionSD2022} Sophisticated analysis of multidimensional, low-noise data is a prerequisite for unambiguous identification of subtle yet important effects in photochemical dynamics, including interferences or passage through conical intersections, and inversion methods are expected to play an increasingly central role in ultrafast imaging.

\begin{acknowledgement}
KA acknowledges an EPSRC doctoral studentship and a public engagement scholarship from the University of Edinburgh. AK acknowledges funding from the EPSRC (EP/V006819 and EP/V049240) and the Leverhulme Trust (RPG-2020-208). This work was also supported by the Department of Energy, Office of Science, Basic Energy Sciences, under award number DE-SC0020276. Finally, AK acknowledges a Fellowship at the Swedish Collegium for Advanced Studies with financial support from the Erling-Persson Family Foundation and the Knut and Alice Wallenberg Foundation. 
\end{acknowledgement}

\begin{suppinfo}
Further details are provided in the Supplementary Information.
\end{suppinfo}

\bibliography{jctc_refs}

\end{document}

\newpage

\section{Limitations of $\Delta\mathrm{PDF}(R,t)$}\label{sec:pdf_limitations}

\begin{figure}[hbt!]
    \centering
     \captionsetup{justification=centering,margin=6em}
    \includegraphics[width=4in, height=6.5in, scale=0.5]{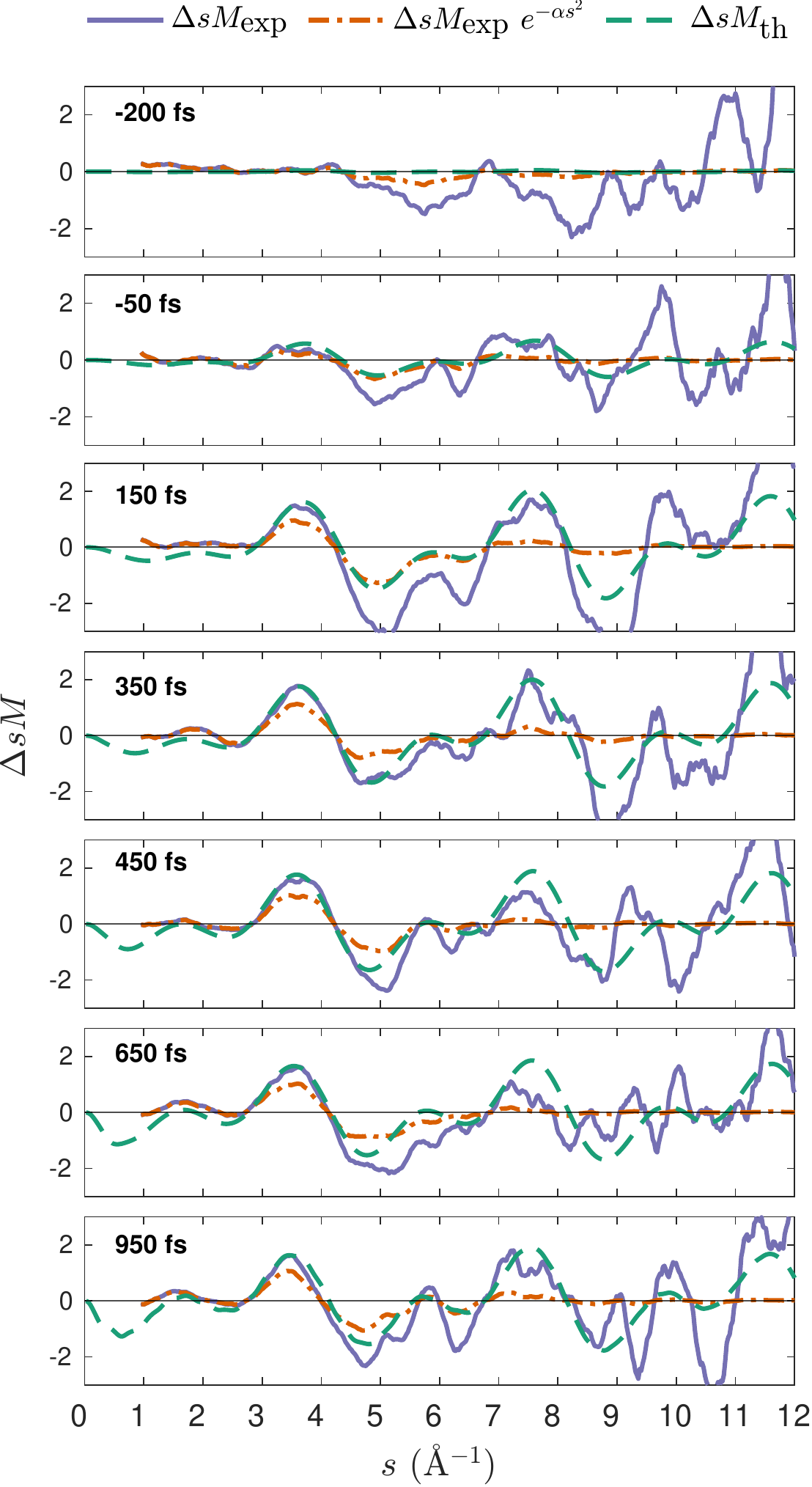}
    \caption{Comparison of experimental and theoretical $\Delta sM$ signal as a function of time. Note the inclusion of the dampened experimental $\Delta sM e^{-\alpha s^2}$ which is transformed into real space to obtain the pair distribution function. The noise at high $s$ is heavily reduced at the risk of reducing the intensity of the signal in the intermediate range. Note the absence of experimental data at low values of $s$. }
    \label{fig:dsm_compare}
\end{figure}

It is common to represent time-dependent electron scattering signals in terms of the modified scattering $\Delta\mathrm{sM}(s,t)$, defined as,

\begin{equation}
    \Delta\mathrm{sM}(s,t) \approx 
    s\frac{I_{\mathrm{on}}(s,t)-I(s)_{\mathrm{off}}}{I_{\mathrm{at}(s)}}.
    \label{eqn:DeltasM}
\end{equation}
Here we use the electron scattering convention of denoting the momentum transfer vector as $s$. $I_{\mathrm{on}}(s,t)$ and $I_{\mathrm{off}}(s)$ refer to the signal at time $t$ and the $t\ll0$ signal recorded without a pump pulse. In the denominator, we find the background that results from the incoherent summation of the scattered intensity from individual atoms, which is denoted $I_{\mathrm{at}}(s)$. While the signal represented by defined eq~\ref{eqn:DeltasM} allows one to sine-transform the signal into real space, the division by the {\it{theoretically}} calculated $I_{\mathrm{at}}(s)$ also results in an absence of error cancellation in comparison to the signal represented as $\% \Delta I_{\textrm{exp}}(s,t)$. See Figure~\ref{fig:dsm_compare} for lineouts of the $\Delta sM$ signal. Thus, we note that for data with poor signal-to-noise ratio, it is often beneficial to perform the inversion procedure in the reciprocal space of $\% \Delta I_{\textrm{exp}}(s,t)$. 

\begin{figure}[hbt!]
    \centering
    \captionsetup{justification=centering,margin=6em}
    \includegraphics[width=4in, height=6.5in, scale=0.5]{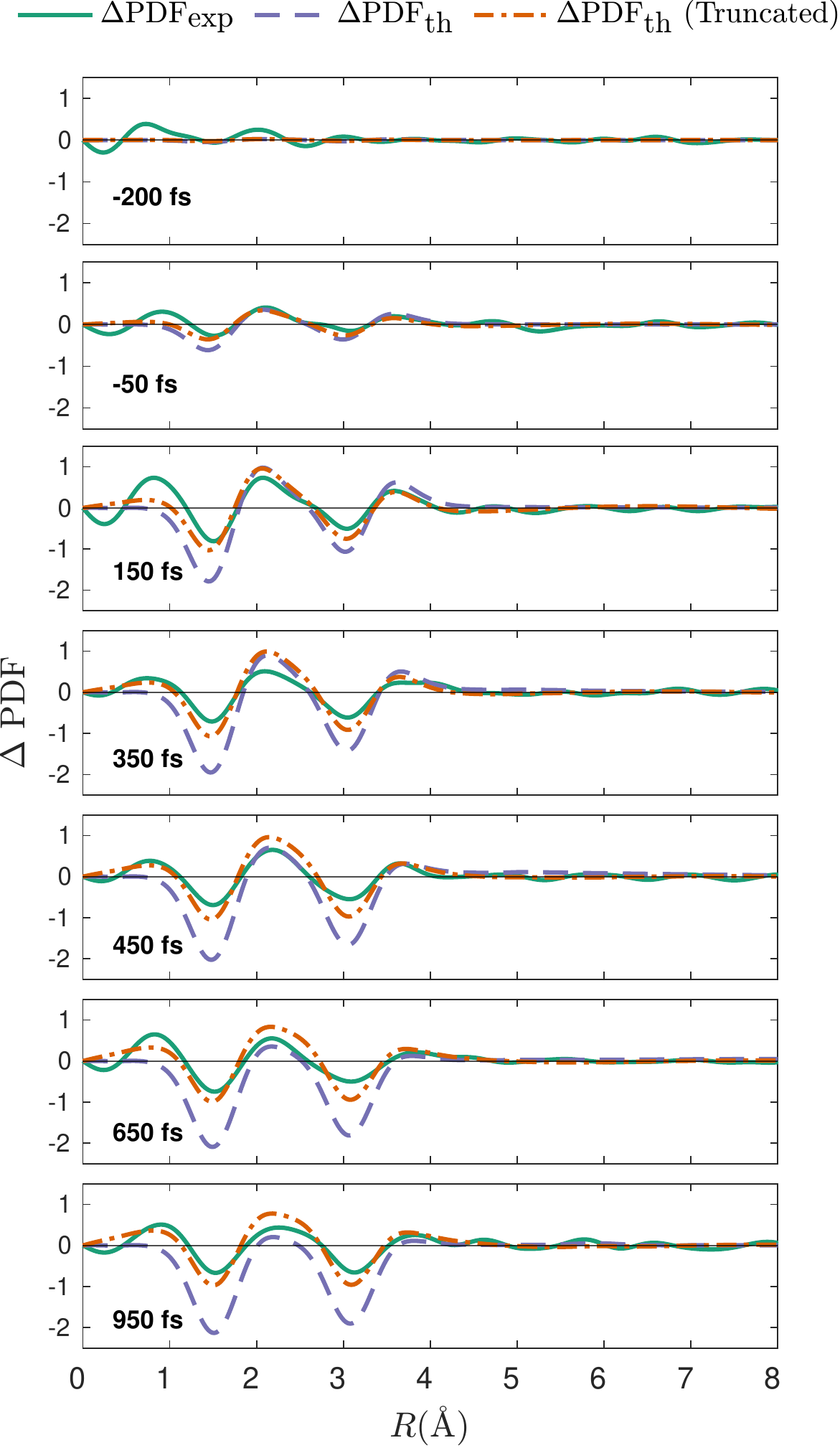}
    \caption{Lineouts of experimental and theoretical $\Delta PDF$. The experimental $\Delta PDF_\mathrm{exp}$ and the theoretical $\Delta PDF_\mathrm{th}$ (truncated), which results from truncating the transformation range to exclude the low $s$ region, show good agreement. The absence of the low frequencies from small $s$ yields a vertical shift in the $\Delta PDF$s and an unphysical shoulder at low atomic separations. As a consequence, these two curves indicate the presence of a very short and an intermediate ($\approx2.5$~\AA) bond length that have no physical basis in \ce{CS2}, especially not at large times $t$. In contrast, the $\Delta PDF$ calculated from the theoretical $\Delta sM$ which includes the low $s$ data does not suffer these artifacts.}
    \label{fig:pdf_lineouts}
\end{figure}

The sine-transform that allows for the direct inversion of $\Delta\mathrm{sM}(s,t)$ to real space can be defined as,
\begin{equation}
    \Delta\mathrm{PDF}(R,t)
    =
    \int_{s_{\mathrm{min}}}^{s_{\mathrm{max}}}\Delta \mathrm{sM}(s,t)\sin(sR)\mathrm{e}^{-\alpha s^{2}} ds.
    \label{eqn:DeltaPDF}
\end{equation}
The resulting pair distribution function, $\Delta\mathrm{PDF}(R,t)$, reveals the change in the distribution of interatomic distances $R$ with time. Ideally, the limits of integration in eq~\ref{eqn:DeltaPDF} should extend from $s_{\mathrm{min}} = 0$ to $s_{\mathrm{max}} = \infty$. In reality, for current MeV-UED experiments, one can expect an upper limit $s_{\mathrm{max}} \approx 12$ \AA. Moreover, the reliability of the higher range of $s$ values is limited due to the $s^{-4}$,\cite{Weber2020} hence a dampening factor of $\mathrm{e}^{-\alpha s^{2}}$ is included to limit the effect of large $s$ noise on the transform. Care must be taken in the tuning of the value of $\alpha$, there is a balance between limiting the detrimental effect of noise at high $s$ and ensuring that the range of $s$ included is sufficiently large to accurately reproduce the distribution of interatomic distances. At the same time, the signal is usually not recorded in the very low range of $s$ due to centre-shifting and streaking resulting from plasma effects in the target. In the event that only a small amount of data is missing from the low range, linear extrapolation to $s = 0$ can be attempted. However, when $s_\mathrm{min}$ is too large, this can result in inaccurate or erroneous PDFs. In brief, the lack of data at low $s$ amounts to the exclusion of low frequencies in the sine transform, resulting in the PDF being artificially shifted by a near constant term. The experimentally determined PDF will then exhibit interatomic distances below that of the acceptable physical limit, as well as an unexpected increase in intermediate distances at later times. If the size of the transform window is truncated to exclude the region below 1~\AA, then the theoretical PDF can be made to match the experimentally determined PDF, as in Figure~\ref{fig:pdf_lineouts}. While one could envision performing our procedure directly in real space for very high quality UED data in the future, due to these reasons, we perform the fitting procedure in reciprocal space on the signal represented as $\% \Delta I_{\textrm{exp}}(s,t)$. Overall, this is a much more stable alternative.

\section{Iterative Weight Generation Procedure}\label{sec:weights}
In cases where convergence is not easily achieved through the standard Monte-Carlo approach to weight generation on the interval [0, 1], it may be desirable to sample the total weight configuration space iteratively through a series of configuration space samples increasing in size. This involves defining a series of bounding limits $\left( L_{\mathrm{bound}}, U_{\mathrm{bound}} \right)$ on the allowed values of the weights. One starts the procedure with a set of bounds that highly constrain the sample space so that the generated initial weights are narrowly distributed around that of the equally weighted case, i.e.\ where $\mathbf{w} = 1/N_{\textrm{TBF}}$. The target function is then minimised with respect to each of the $N_{\mathrm{init}}$ initial weights in the pool. If required, the number of sets of initial weights $N_{\mathrm{init}}$ can be selectively tuned so that convergence of the target function to the minima within the allowed bounds is observed. From this initial sample of the weight space, the sets of initial conditions that give rise to the lowest value of the target function are selected and entered into the subsequent pool of initial conditions that are generated by slightly increasing the bounds $\left( L_{\mathrm{bound}}, U_{\mathrm{bound}} \right)$ on the allowed values of the weights. This weight sampling procedure is then repeated for a series of bounds, each of which correspond to relieving some of the constraint on the allowed values and hence amount to increasing the size of the weight configuration space which is sampled. At each step, the best set of initial conditions are selected and inserted into the next iterations pool. One must be careful that the number of initial conditions $N_{\mathrm{init}}$ at each iteration is sufficient to allow convergence. Generally, $N$ should increase as the the bounds are lifted as the size of the weight sampling space is increasing. This is repeated until the bounds are such that the weights are sampled on the interval [0, 1], which corresponds to the whole weight configuration space. By projecting through a series of different sized weight sampling spaces we provide a way of ensuring the sampling of the space is sufficient enough for convergence to correct global minima to be achieved. It also allows for the refinement of the number of sample sets $N_{\mathrm{init}}$ so that it is large enough for global convergence without having to blindly select a huge value which may result in a large computational overhead.

\section{Data Treatment}\label{sec:data_treatment}
\subsection{Ultrafast x-ray scattering (UXS)}\label{sec:uxs_treatment}
\begin{figure}[hbt!]
    \centering
    \captionsetup{justification=centering,margin=6em}
    \includegraphics[width=4in, scale=0.75]{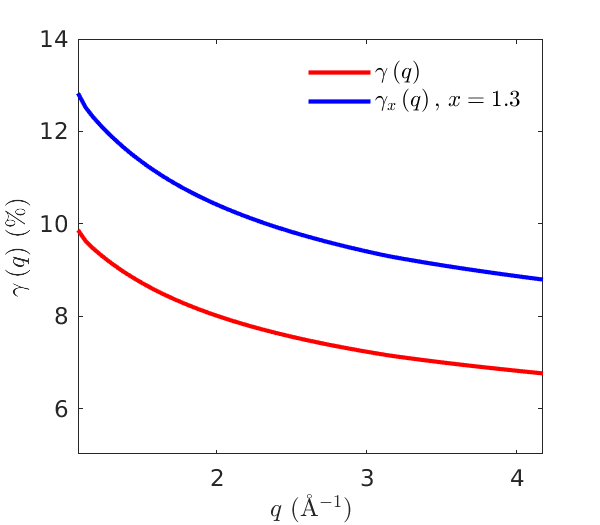}
    \caption{The $q$ dependent excitation fraction $\gamma\left(q\right)$ used in the CHD optimisation. This mitigates the effect of different contributions of $\gamma\left(q\right)$ across $q$ due to the large interaction region. This is scaled uniformly across $q$ in the target function.}
    \label{fig:chd_xfrac}
\end{figure}

Data was recorded over the time range $-1 \leq t_e \leq 4$~ps and ranged from $0.925 \leq q \leq 4.175$ \AA$^{-1}$.\cite{minitti_imaging_2015} An initial value of $t^{\prime} = 425$ fs was selected as an approximate time-zero. In the UXS case, we scale the signal according to fluctuations in the beam intensity. With this in mind, the generalised percent difference signal becomes,
\begin{equation} \label{eq:Idiff_xray}
\%\Delta I_{\textrm{mod}}(q,t,\gamma)
  =
  \frac{100}{1+\beta(t,\gamma)}\left(\gamma(q) \left[ \frac{I_{\textrm{on}}(q,t)}{I_{\textrm{off}}(q)}-1 \right] -\beta(t,\gamma)\right),
\end{equation}
where $\beta(t,\gamma)$ is given by,
\begin{equation}
    \beta(t,\gamma) = \frac{Q^{\gamma}(t)}{Q_{\textrm{off}}}.
\end{equation}
Here $Q_{\textrm{off}}$ and $Q^{\gamma}(t)$ refer to the integrated signal on the detector with the laser off and on respectively. Notice the excitation fraction $\gamma\left(q\right)$ now has a dependence on $q$, this can be seen in Figure~\ref{fig:chd_xfrac}. The purpose of the explicit dependence of $\gamma\left(q\right)$ on $q$ is to account for the non-uniform nature of the scattered intensity across the measured $q$ range, this arises from the long and extended interaction region. Note, we do not optimise $\gamma\left(q\right)$ directly with respect to $q$. Rather, we optimise a scaling factor $x$ that acts to uniformly scale $\gamma\left(q\right)$ across the available $q$ range such that it yields a scaled optimised excitation fraction $\gamma_{x}\left(q\right)$, which is related to the initial estimate of the $q$ dependent excitation fraction $\gamma\left(q\right)$ as, $\gamma_{x}\left(q\right) = \gamma\left(q\right) x$. The initial guess of $\gamma\left(q\right)$ is calculated from Beer-Lambert law. Clearly $Q^{\gamma}(t)$ is excitation fraction and $q$ dependant. The laser-on case is defined as,
\begin{equation}
    Q^{\gamma}(t) = 2\pi \int_{q_0}^{q_1} F_{\textrm{instr}}(q)\gamma(q) \left[ I_{\textrm{on}}(q,t) - I_{\textrm{off}}(q) \right]q dq,
\end{equation}
where $F_{\textrm{instr}}(q)$ is the instrument response function that accounts for variation in detector sensitivity with $q$. The limits of integration $q_0$ and $q_1$ refer to the radial limits on the detector. The laser off intensity is given as,
\begin{equation}
    Q_{\mathrm{off}} = 2\pi \int_{q_0}^{q_1} F_{\textrm{instr}}(q) I_{\textrm{off}}\left(q\right) q dq.
\end{equation}
In the event that the intensity is equal for both the on and off pulses, $\beta$ reduces to zero and eq~\ref{eq:Idiff_xray} reduces to the generalised expression in the main manuscript. As outlined in the main text, the percent difference signal is calculated from a series of trajectories assuming the IAM and using the x-ray scattering form factors $f_{A}^{x}\left(q\right)$.\cite{IntTabCryVolC} In the case of UXS, the beam is linearly polarised, and so we scale $\%\Delta I_{\mathrm{mod}}(q,t,\gamma)$ by a polarisation factor $\frac{1}{2}\left(1+\cos^2{\theta}\right)$. Next, the signal $\%\Delta I_{\mathrm{mod}}(q,t,\gamma)$ is convoluted to mimic the temporal resolution observed in experiment, as described in the main manuscript. The duration of the UV pump laser and the x-ray pulse were measured as 60~fs and 30~fs respectively, thus the value of $\tau_c$ is fixed to the product of two Gaussian's with the corresponding width as to mimic the instrument response function. Finally, the signal is binned along the temporal axis into bins of size $\Delta t = 25$~fs. Accounting for all of this, the length of the subset of the signal used in the optimisation becomes $t^{\prime} + 275$~fs.

To limit the effect of data points with lower statistics on the minimisation of $F\left(\mathbf{w}, \mathbf{c}\right)$, we define the confidence matrix $p_\mathrm{conf}$ based on the number of photon hits per frame as,
\begin{equation}\label{eqn:conf_mat_xray}
p_\mathrm{conf}\left(q_i,t_j^\prime\right) = \frac{N\left(t_j^\prime\right)}{N_{\textrm{max}}} p_{q}\left(q_j\right),
\end{equation}
where $N(t_j^\prime)$ and $N_{\textrm{max}}$ are the number of hits per frame and the maximum number of hits over all frames respectively. The relative accuracy of each point in $q$ is given by $p_{q}(q_j)$.

\subsection{Ultrafast electron diffraction (UED)}\label{sec:ued_treatment}
Data was recorded over the range of momentum transfer $0.76 \leq s \leq 12.25$ \AA$^{-1}$ and over a temporal range of $-1.26 \leq t_e \leq 1.73$ ps.\cite{razmus_multichannel_2022} An initial value of $t' = -120$~fs was selected. In the \ce{CS2} UED case, we define the percentage difference signal in a somewhat simpler fashion as,
\begin{equation} \label{eq:Idiff_ued}
\%\Delta I_{\mathrm{mod}}(s,t,\gamma)
  =
  100\gamma\times
  \frac{I_{\mathrm{on}}(s,t) - I_{\mathrm{off}}(s)}{I_{\mathrm{off}}(s)}.
\end{equation}
 We use relativistic electron scattering form factors $f_{A}^{e}\left(s\right)$ calculated using ELSEPA.\cite{salvat_elsepadirac_2005} 
 
 Note that due to instabilities in the \textit{ab-initio} electronic structure calculations at very large nuclear separations, some dissociative trajectories crash before they reach $t=1$~ps. These trajectorties are extended beyond the point of crashing using a harmonic model of the vibrational motion of the \ce{CS} fragment remaining. The dissociated atom is extended linearly according to its last known velocity. We define dissociation as the point at which a trajectory reaches a \ce{C-S} bond length greater than 3.4~\AA, which is the point of no return in the simulations. 
 
\begin{figure}[hbt!]
    \centering
    \captionsetup{justification=centering,margin=6em}
    \includegraphics[width=4in, scale=0.75]{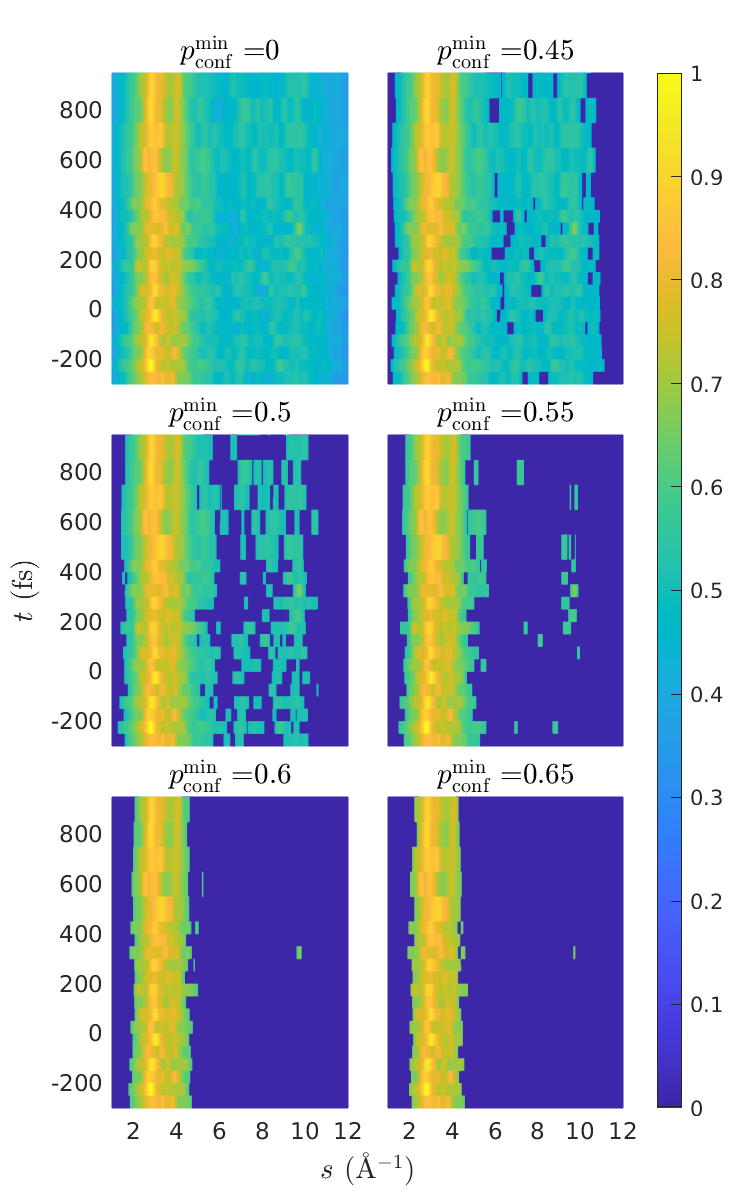}
    \caption{The confidence matrices $p_\mathrm{conf}(q_i,t_j^\prime)$ for a series of thresholds $p_{\mathrm{conf}}^{\mathrm{min}}$ used in the optimisation process of \ce{CS2}. Each point below the chosen value of $p_{\mathrm{conf}}^{\mathrm{min}}$ corresponds to thresholding out the data by setting to zero.}
    \label{fig:cs2_confmats}
\end{figure}

 Due to a lack of statistics on the electron beam, we define a confidence matrix $p_\mathrm{conf}$ as heuristically based on estimated experimental standard deviations from a bootstrapping procedure\cite{razmus_multichannel_2022}, $\sigma(q_i,q_j^\prime)$. The elements of the confidence matrix are normalised as follows,
\begin{equation} \label{eq:conf_mat_ued}
p_\mathrm{conf}(q_i,t_j^\prime) = \left(\frac{\sigma(q_i,t_j^\prime)}{\min{\{\sigma(q_i,t_j^\prime)\}}}\right)^{-1}.
\end{equation}
The matrices $p_\mathrm{conf}(q_i,t_j^\prime)$ that result from setting $p_\mathrm{conf}(q_i,t_j^\prime) \leq p_{\mathrm{conf}}^{\mathrm{min}}$ to zero, where $p_{\mathrm{conf}}^{\mathrm{min}} \in \left[0, 0.45, 0.50, 0.55, 0.60, 0.65 \right]$ can be visualised in Figure~\ref{fig:cs2_confmats}. The case where $p_{\mathrm{conf}}^{\mathrm{min}} = 0$ amounts to including all the data points, but weighting them according to the values in $p_\mathrm{conf}(q_i,t_j^\prime)$. The higher the value of $p_{\mathrm{conf}}^{\mathrm{min}}$, the more data points are excluded from the optimisation. The largest value of $p_{\mathrm{conf}}^{\mathrm{min}} = 0.65$ amounts to including only a narrow band of the main enhancement seen in the signal.

%As a means of testing convergence, we also perform the optimisation procedure using the weight generation procedure discussed in Section~\ref{sec:weights}. We define a set of 35 upper bounding limits $U_{\mathrm{bound}}$, where $U_{\mathrm{bound}} \in [0.075, 1]$. Note that the set of limits is non-linear and the corresponding values can be seen in Figure~\ref{fig:cs2_weight_grid}. For each $U_{\mathrm{bound}}$ we perform a constrained optimisation starting with generating $N = 100$ initial values of $\mathbf{w}$. At each successive iteration we take the best three sets of weights from the previous iteration and add them to the pool of new initial conditions.

\section{Additional Comments on Convergence}\label{sec:convergence}

\subsection{Initial Conditions}\label{sec:condition_convergence}
\begin{figure}[hbt!]
    \centering
    \captionsetup{justification=centering,margin=6em}
    \includegraphics[width=4in, scale=1]{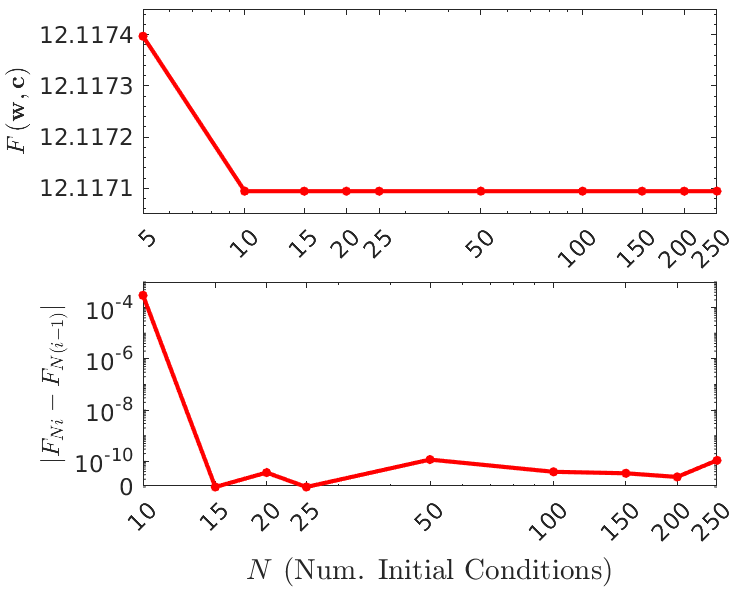}
    \caption{Convergence with respect to the number of initial conditions $N_{\mathrm{init}}$ in the Monte-Carlo sampling procedure for the \ce{CS2} reaction. The figure shows the convergence of the minimum value of the target function $F\left(\mathbf{w}, \mathbf{c}\right)$ \textit{(top panel)} and also the difference between $F\left(\mathbf{w}, \mathbf{c}\right)$ at each successive sampling density governed by $N_{\mathrm{init}}$ \textit{(bottom panel)}. Both are plotted on a logarithmic scale.}
    \label{fig:init_conds_conv}
\end{figure}

Given the Monte-Carlo nature of the sampling of initial trajectory weights $\mathbf{w}$, the choice in the number of initial conditions $N_{\mathrm{init}}$ is crucial. Too small a number of $N_{\mathrm{init}}$ risks that the sampling density of the weight configuration space is inadequate for ensuring that the global minimum is found. In Figure~\ref{fig:init_conds_conv}, we see that the target function converges rapidly as a function of the number of initial conditions $N_{\mathrm{init}}$ used for the optimisations. We observe the lowest value of $F\left(\mathbf{w}, \mathbf{c}\right)$ when $N_{\mathrm{init}} = 250$. However, from $N_{\mathrm{init}} = 15$ onwards, we see that $F\left(\mathbf{w}, \mathbf{c}\right)$ only changes by $1e{-10}$ between each successive increase in $N_{\mathrm{init}}$. For $N_{\mathrm{init}} = 100$, which is the value of $N_{\mathrm{init}}$ we use in unconstrained optimisations, we see that 46~\% of the initial conditions within the sample pool converge to a solution within $1e{-3}$ of the best optimisation.

\begin{figure}[hbt!]
    \centering
    \captionsetup{justification=centering,margin=6em}
    \includegraphics[width=4in, scale=1]{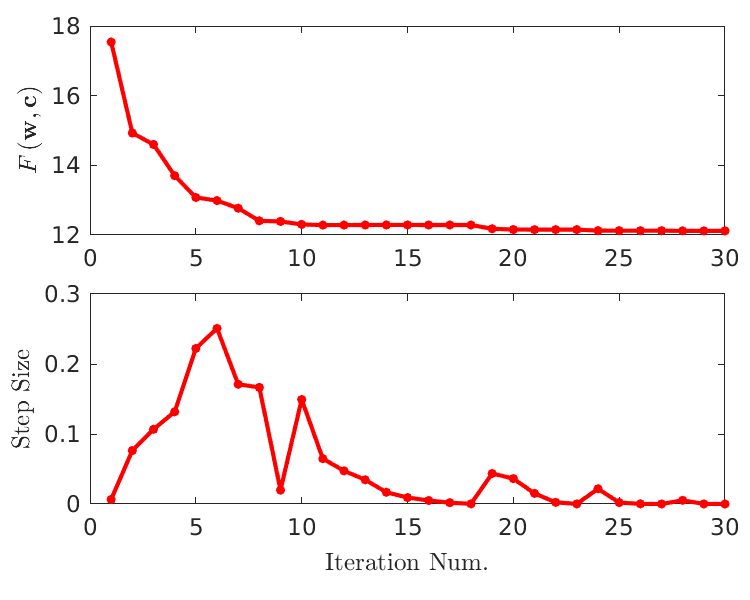}
    \caption{The convergence of the best selected optimisation for \ce{CS2}, with respect to the number of iterations. The \textit{top panel} shows the convergence of $F\left(\mathbf{w}, \mathbf{c}\right)$, and the \textit{bottom panel} shows the convergence of the step size taken along the target function surface at each iteration in the optimisation.}
    \label{fig:best_opt_conver}
\end{figure}

For the best optimisation, the convergence of the value of the target function and the step size at each iteration in the minimisation, can be seen in Figure~\ref{fig:best_opt_conver}. The value of $F\left(\mathbf{w}, \mathbf{c}\right)$ decreases rapidly within the first few iterations as the step size increases. By iteration 19, $F\left(\mathbf{w}, \mathbf{c}\right)$ reaches a value of 12.18, before converging to 12.12 as the size of the step decreases and the global minimum is located.

\subsection{Iterative Weight Sampling - \ce{CS2}}\label{sec:cs2_iterative_weights}

\begin{figure}[hbt!]
    \centering
    \captionsetup{justification=centering,margin=6em}
    \includegraphics[width=4in, scale=0.75]{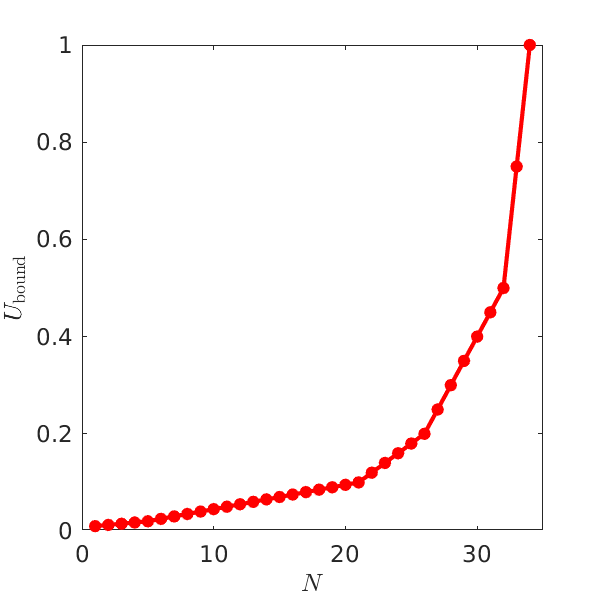}
    \caption{The grid of 35 upper bounds $U_\mathrm{bound}$ for which a series of constrained optimisations are performed. The end case where $U_\mathrm{bound} = 1$ equates to the unconstrained optimisation where $\mathbf{w} \in [0, 1]$. }
    \label{fig:cs2_weight_grid}
\end{figure}

Due to the flatness of the target function and the presence of many local minima, we determine that the best optimisation $(t_0, \tau_c, \gamma, p_\mathrm{conf}^\mathrm{min})$ = (83~fs, 230~fs, 3.4\%, 0) is the true global minimum through the weight generation procedure outlined in Section~\ref{sec:weights}. We define a set of 35 upper bounding limits $U_{\mathrm{bound}}$, where $U_{\mathrm{bound}} \in [0.075, 1]$. Note that the set of limits is non-linear and the corresponding values can be seen in Figure~\ref{fig:cs2_weight_grid}. For each $U_{\mathrm{bound}}$ we perform a constrained optimisation that starts by generating $N_{\mathrm{init}} = 100$ initial values of $\mathbf{w}$. At each successive iteration we take the best three sets of weights from the previous iteration and add them to the pool of new initial conditions.

\begin{figure}[hbt!]
     \centering
     \includegraphics[width=4in]{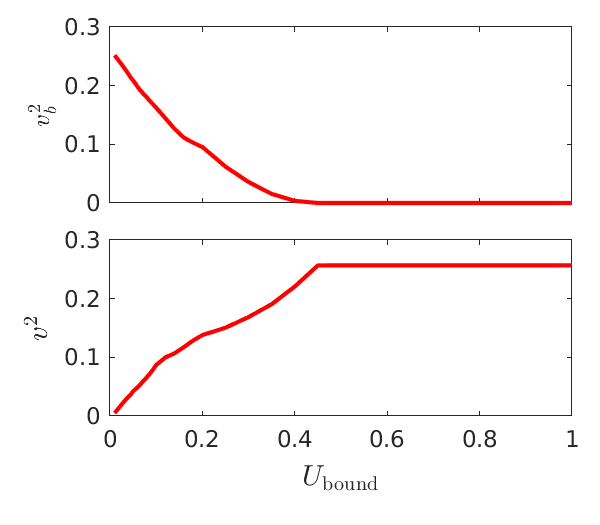}
     \caption{Variance of the final weights for a series of different $U_{\mathrm{bound}}$ values from the best set of weights (top panel) and the variance from the mean weight in each optimisation (bottom panel) i.e. the spread of the weights.}
     \label{fig:var_weight_space}
 \end{figure}

\begin{figure}[hbt!]
    \centering
    \captionsetup{justification=centering,margin=6em}
    \includegraphics[width=4in]{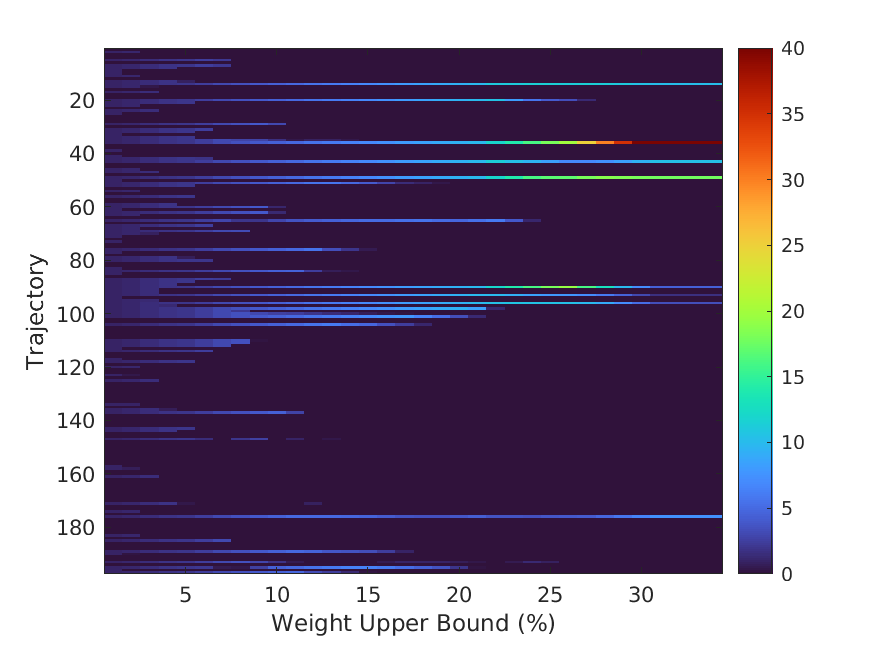}
    \caption{Heatmap of the individial trajectory weights as a function of the upper bound $U_{\mathrm{bound}}$ varied in the series of constrained optimisations for \ce{CS2}. Highly constrained optimisations result in the weight being distributed more evenly between a broader range of trajectories. As constraint is lifted, the archetypical trajectories are filtered out.}
    \label{fig:constrained_traj_weights}
\end{figure}

 Generally, the same set of trajectories are picked out in all the optimisations, although the lower the value of $U_{\mathrm{bound}}$, the more trajectories are given an appreciable weight. However, these trajectories are very narrowly distributed. This is apparent in the lower panel of Figure~\ref{fig:var_weight_space}, where one can see how small the variance $v^{2}$ is for the set of weights that result in the lowest values of $F\left(\mathbf{w}, \mathbf{c}\right)$ corresponding to local minima in close proximity to the global minimum. This can be explained as by forcing a broader distribution in the number of trajectories selected through placing significant constraint on the optimisation, the trajectories are all given a weight close to the limit of $U_{\mathrm{bound}}$. As the constraint is lifted, the variance within the weights $\mathbf{w}$ increases. This effect can also be seen in Figure~\ref{fig:constrained_traj_weights}, which shows the trajectory weights as a function of $U_{\mathrm{bound}}$. Once the value of $U_{\mathrm{bound}} = 0.45$ is reached, the weights converge and the optimisation distills out the key dynamics from the broad distribution of trajectories. This is also reflected in the upper panel of Figure~\ref{fig:var_weight_space}, which shows the variance of the weights for each $U_{\mathrm{bound}}$ from the best optimisation where $(t_0, \tau_c, \gamma, p_\mathrm{conf}^\mathrm{min}) =$ (83~fs, 230~fs, 3.4\%, 0).
 
\begin{figure}[hbt!]
    \centering
    \includegraphics[width=4in]{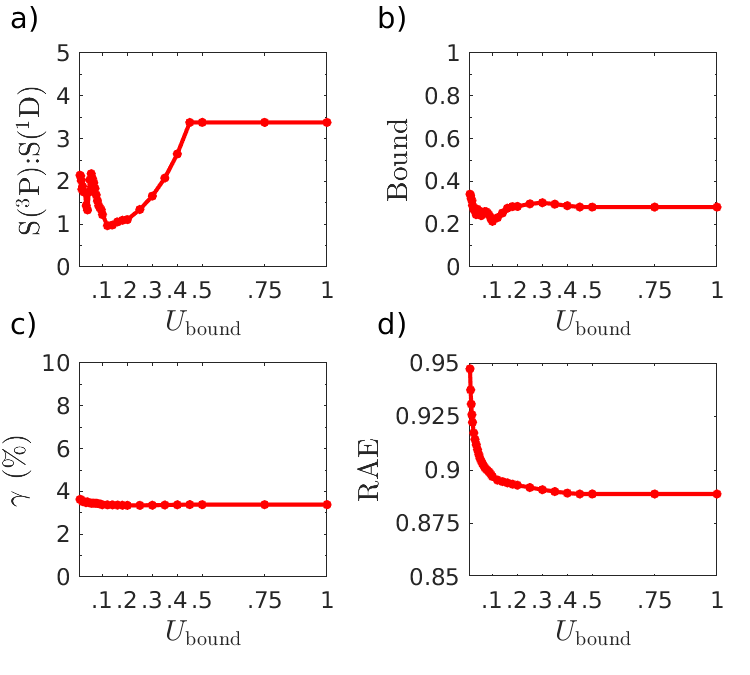}
    \caption{The convergence of a) branching ratio, b) bound fraction at 1 ps, c) excitation fraction $(\gamma)$, and the relative absolute error as a function of the upper bound on the size of the weight space from which the distribution of initial conditions are generated.}
    \label{fig:weight_space_conv}
\end{figure}

 Figure~\ref{fig:weight_space_conv} shows the convergence of key parameters as a function of $U_{\mathrm{bound}}$. The lack of sensitivity to the value of $\gamma$ is apparent, indicating that the initial guess of $\gamma$ as determined in the initial $t_0$ fit is reasonable. The convergence of the bound fraction appears to take a bit longer, although values remain in the range of 20-30\% for the most part. In comparison, the convergence of the branching ratio is significantly slower, the predicted ratio starting off around 1:2 which is in reasonable agreement with literature. However, this quickly gets worse as some constraint on the optimisation are lifted. The effect is due to over-estimation of the contribution of the singlet population, and it is not until there is sufficient freedom that some of this weight is redistribution to several triplet trajectories. The failure of the constrained optimisations to pick out the finer details is not surprising, given the low resolution in the experimental data and that there is some linear dependence between the singlet and triplet contributions to dissociation.
 
\newpage

\newpage

\newpage

\begin{figure}[hbt!]
    \centering
    \captionsetup{justification=centering,margin=6em}
    \includegraphics{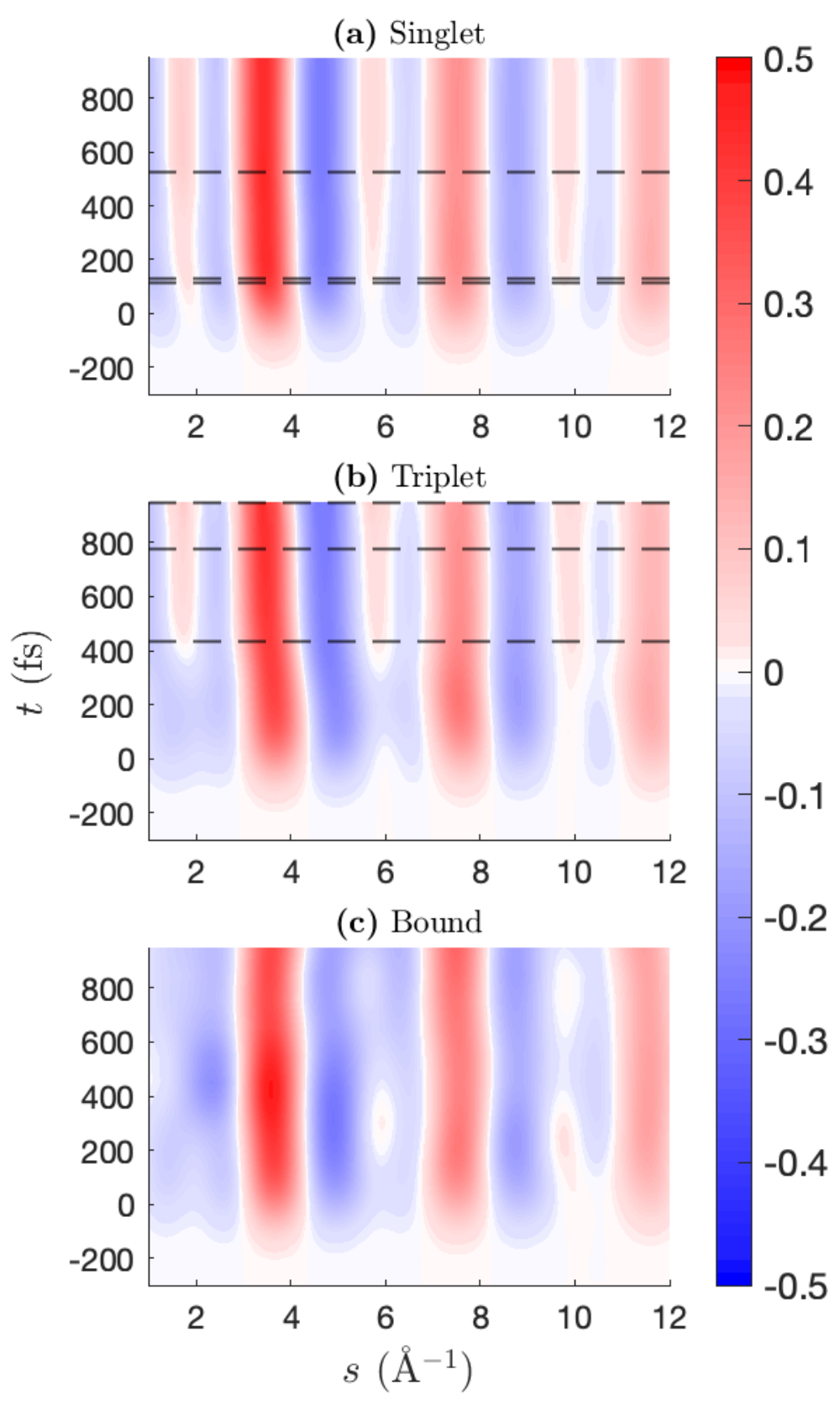}
    \caption{The optimised model \ce{CS2} signal $\%\Delta I_{\mathrm{mod}}\left(s, t\right)$ decomposed into its contributions from \textit{a) dissociative singlet}, \textit{b) dissociative triplet}, and \textit{c) bound trajectories}. Dashed black lines represent the dissociation times of the individual trajectories from which the signal is composed. Note the presence of the peak just below 2~\AA$^{-1}$ that is delayed in onset and the shift in the main enhancement, these are both correlated with dissociation.}
    \label{fig:cs2_signal_decomp}
\end{figure}

\newpage

\begin{figure}[hbt!]
    \centering
    \captionsetup{justification=centering,margin=6em}
    \includegraphics[width=4in, scale=0.75]{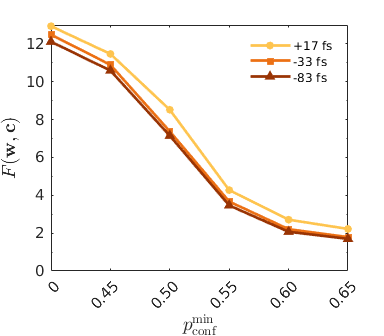}
    \caption{The value of $F\left(\mathbf{w}, \mathbf{c}\right)$ with respect to confidence threshold $p_{\mathrm{conf}}^{\mathrm{min}}$ at three $t_0$ shifts. Note that comparisons between different values of $p_{\mathrm{conf}}^{\mathrm{min}}$ hold no meaning as thresholding data points will always lead to a lower value of $F\left(\mathbf{w}, \mathbf{c}\right)$. The general trend that $t_0 = -83$~fs gives the lowest value is true across all values of $p_{\mathrm{conf}}^{\mathrm{min}}$. This includes $p_{\mathrm{conf}}^{\mathrm{min}} = 0$ which is selected as optimal when the convergence of the statistical measures and physical constants are taken into account, as discussed in the main manuscript.}
    \label{fig:cs2_tfunc}
\end{figure}

\newpage

\begin{figure}[hbt!]
    \centering
    \captionsetup{justification=centering,margin=6em}
    \includegraphics[width=4in, scale=0.75]{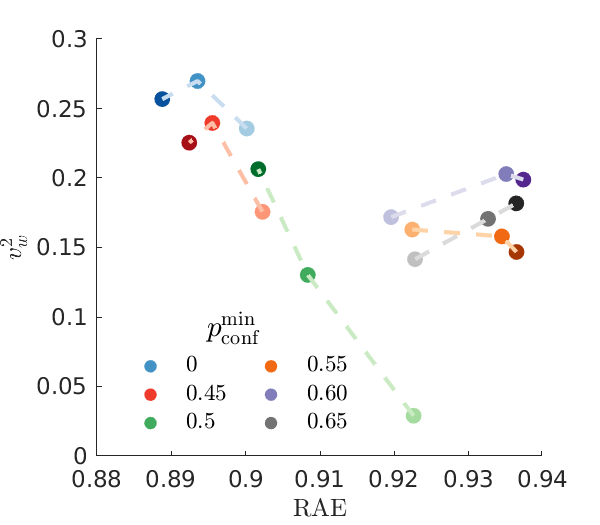}
    \caption{The variance $v_{w}^{2}$ of the best set of weights for a series of \ce{CS2} optimisations for which $p_{\mathrm{conf}^{\mathrm{min}}}$ and $t_0$ are varied. For each value of $p_{\mathrm{conf}}^{\mathrm{min}}$, the varying opacity represents a different $t_0$ shift, ranging from -83, -33 to +17~fs from dark to light respectively. The variance of the weights increases for the better optimisations with a lower relative absolute error (RAE).}
    \label{fig:cs2_var_weights}
\end{figure}

\newpage

\begin{figure}[hbt!]
    \centering
    \captionsetup{justification=centering,margin=6em}
    \includegraphics[width=4in, scale=0.75]{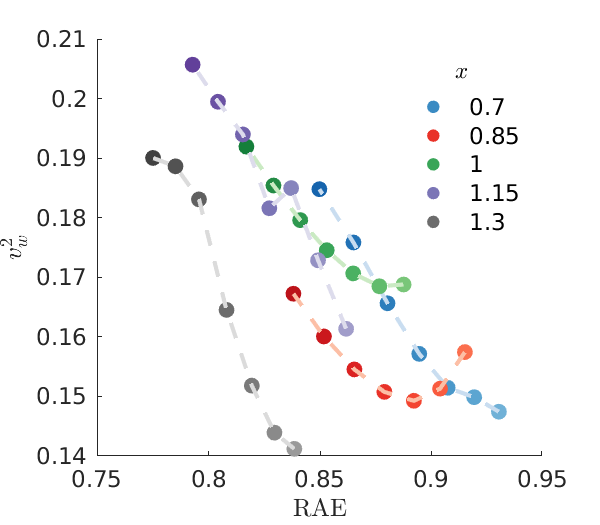}
    \caption{The variance $v_{w}^{2}$ of the best set of weights for a series of CHD optimisations for which the scaling factor $x$ and $t_0$ are varied. For each value of $x$, the varying opacity represents a different $t_0$ ranging from -38 to -14~fs (the darker the shade, the earlier the $t_0$ shift). The variance of the weights tends to increase for the better optimisations with a lower relative absolute error (RAE).}
    \label{fig:chd_var_weights}
\end{figure}

\newpage

\newpage

\newpage

\newpage

\newpage

 \newpage

\newpage

\newpage

\newpage

\bibliography{jctc_refs}